\documentclass[review,3p,times]{elsarticle}

\usepackage{amssymb}
\usepackage{amsmath}
\usepackage{siunitx}
\usepackage{xfrac}
\usepackage{xparse}
\usepackage{xcolor}
\usepackage{longtable,tabularx}
\usepackage{multirow}
\usepackage{makecell}
\usepackage{rotating}
\usepackage{soul}
\usepackage{bm}
\usepackage{booktabs}
\usepackage{algorithmicx}
\usepackage{algorithm}
\usepackage{algpseudocode}
    
\usepackage{etoolbox}
\usepackage{tikz}
\usetikzlibrary{tikzmark}
\usetikzlibrary{calc}

\usepackage[utf8]{inputenc} 
\DeclareUnicodeCharacter{2002}{\hspace{0.5em}} 

\algrenewcommand\algorithmicrequire{\textbf{Input:}}
\algrenewcommand\algorithmicensure{\textbf{Output:}}
\algrenewcommand{\algorithmiccomment}[1]{%
      \hfill\mbox{$\triangleright$~#1}%
}

\errorcontextlines\maxdimen

\makeatletter
\newcommand{\multiline}[1]{%
      \begin{tabularx}{\dimexpr\linewidth-\ALG@thistlm}[t]{@{}X@{}}
            #1
     \end{tabularx}
}
\makeatother

\algnewcommand\True{\textbf{true}\space}
\algnewcommand\False{\textbf{false}\space}

\newcommand{\norm}[1]{\left\lVert#1\right\rVert}

\renewcommand{\figurename}{Fig.}
\newcommand{\sectionname}{Sec.}

\journal{Computer Methods in Applied Mechanics and Engineering}

\begin{document}
	
	\begin{frontmatter}
		
		
		
        \title{Adaptive Multi-Fidelity Structural Optimization under Fluid-Structure Interaction}
				
		
		\author[label1,label2]{Aditya Narkhede} 
		\author[label1]{Erick Rivas}
		\author[label1]{Kevin Wang\corref{cor1}}
		
		\affiliation[label1]{
			organization={Kevin T. Crofton Department of Aerospace and Ocean Engineering},
			addressline={Virginia Tech}, 
			city={Blacksburg},
			state={VA},
			country={USA}}
		\affiliation[label2]{
			organization={Department of Mechanical Engineering},
			addressline={University of British Columbia}, 
			city={Vancouver},
			state={BC},
			country={Canada}}		
		\cortext[cor1]{Corresponding author.}
				\ead{kevinw3@vt.edu}

		\begin{abstract}

		The design of structures and vehicles subject to fluid-structure interaction (FSI) often requires high-fidelity coupled analysis. While the design variables pertain to the structure, the computational cost is usually dominated by the fluid solver, making iterative optimization prohibitively expensive. This paper presents an adaptive multi-fidelity optimization method that combines high-fidelity FSI analysis with a lightweight surrogate for fluid-induced loads and a decision model that selects between surrogate and high-fidelity fluid evaluations.  During optimization, results from completed FSI analyses are used to incrementally update a non-intrusive surrogate model based on nearest-neighbor search and radial interpolation. To maintain accuracy across varying geometries, a hybrid Lagrangian-Eulerian mapping function is developed to transfer fluid loads between structural designs. The evolution in surface orientation is handled through decomposition of the traction vectors in local orthonormal bases. An adaptive Gaussian process regression model is employed to predict surrogate error and quantify uncertainty, allowing risk-aware selection of when coupled analysis is required. As design evaluations cluster near the optimal solution, the accuracy of the surrogate model naturally improves, thereby reducing the reliance on high-fidelity fluid evaluations. This optimization framework treats the high-fidelity FSI model as the ground truth, instead of a pre-computed dataset. It requires no offline training, preserves the high-fidelity structural model in all design evaluations, and ensures that the final design is evaluated by high-fidelity FSI analysis. The fundamental idea is justified theoretically using a simplified model problem, which shows that the leading-order error is a monotonically increasing, concave, and bounded function of the fluid added mass. The framework is implemented using open-source software and demonstrated numerically through two benchmark examples. For shape optimization of a flexible panel under shock loading, results show an $80\%$ reduction in computational cost while maintaining accuracy within $2.3\%$ of fully high-fidelity FSI optimization.
		
		\end{abstract}
		
		
%
		
	\end{frontmatter}
	
	
	
\section{Introduction}\label{sec:introduction}

The motion and deformation of structures in fluid environments are inherently coupled with the fluid flow. When a structure is lightweight and flexible, or the fluid has a high density (e.g., liquids), fluid-structure interaction (FSI) can significantly alter the structure's performance and integrity compared to how it would behave in vacuum or a steady flow. The design of structures subject to FSI is often motivated by the pursuit of energy efficiency \cite{motley2009utilizing,kim2017optimization}, maneuverability \cite{kano2025structural,chung2018cfd}, and biomimicry \cite{marsden2014optimization,long2014shape}. Another class of applications involves systems specifically designed to withstand mechanical or thermal loads transmitted through a fluid medium, such as shock-absorbing armor and explosion containment chambers \cite{ma_computational_2022,narkhede_fluid_2025}.

In such cases, predicting structural dynamics requires coupling it with the surrounding fluid dynamics, although the fluid itself may not be the primary concern. Over the past few decades, a major development in computational mechanics has been the integration of computational fluid dynamics (CFD) and computational structural dynamics (CSD) for high-fidelity FSI analysis. Research efforts have been made to reconcile the different (Eulerian and Lagrangian) reference frames used to formulate fluid and structural kinematics \cite{kamrin2012reference,cerquaglia2019fully}, to enforce interface conditions on moving boundaries \cite{wang2011algorithms,kamensky2015immersogeometric}, and to maintain numerical stability and accuracy \cite{cao2018robin,gonzalez2023three}. This body of work has also intersected with research on conjugate heat transfer, allowing both mechanical and thermal exchanges across the material interface~\cite{miller2018efficient,errera2016comparative,gravemeier2022partitioned}.

In this paper, we consider the numerical optimization of structures subject to FSI. The governing partial differential equations (PDEs) of the fluid and structure naturally appear as equality constraints. While coupled CFD--CSD analyses can be used to evaluate these constraints, their computational cost is often prohibitive for the iterative optimization process. To reduce the cost, a conceptually straightforward idea is to treat the coupled analysis as a black box, and approximate its input-output relation using statistical models. This approach, known as surrogate-based analysis and optimization (SBAO) \cite{queipo_surrogate_2005,liu_gaussian_2013}, is non-intrusive in that it does not require modification of the CFD and CSD solvers. However, it is not specifically tailored to FSI problems, and does not distinguish the fluid and structural responses ~\cite{zhang_optimization_2014,yoshimura_topology_2017,wu_optimizing_2017}.


In coupled CFD--CSD analyses, the computational cost of the fluid solver is often far greater than that of the structure. As fluids do not resist shape changes, the region of fluid flow interacting with a solid structure is typically much larger than the structure itself. The widespread use of efficient geometric representations (e.g., beam and shell elements) further reduces the cost on the structural side. Meanwhile, the design variables are often associated with the structure rather than the fluid. This imbalance between computational cost (dominated by the fluid) and optimization ``relevance'' (dominated by the structure) motivates a strategy that may be well-suited to FSI-constrained structural optimization: retaining the high-fidelity CSD model while accelerating the fluid analysis. In this way, the trade-off lies solely in the accuracy of the fluid-induced loads, but not the structural analysis itself.

Different methods have been explored to construct computationally efficient models of fluid dynamics. One approach is to simplify the governing equations by idealizing the flow physics. Examples include potential-flow panel methods, piston theory, and free-field blast wave models (e.g., the Friedlander equation), which  have been integrated with structural dynamics in aeroelasticity and blast response analyses \cite{ribeiro2023free,ortega2019efficient,ma2010failure}. These models are intuitive and do not require large datasets for calibration. Reduced-order models (ROMs) represent another approach in which full-field flow solutions are sought within low-dimensional subspaces constructed from pre-computed CFD results. In most cases, the nonlinear flux functions in the governing equations also need to be approximated through linearization or ``hyperreduction'' to achieve the expected cost reduction~\cite{zahr2015progressive,tezaur_robust_2022}. Some ROMs have been used in FSI analyses as substitutes for the original CFD model, leading to speedups by several orders of magnitude \cite{lieu2006reduced,xiao2016non}. More recently, neural network-based surrogates have gained attention. Autoencoders and decoders comprised of linear operators and nonlinear scalar activation functions have been used to map between high-dimensional flow fields and low-dimensional latent spaces~\cite{guo_convolutional_2016,murata_nonlinear_2020,obiols_cfdnet_2020,kochkov_machine_2021}. In many cases, neural network surrogates are trained on CFD data, and represent input-output mappings rather than the underlying physics. Nonetheless, physical constraints can be incorporated into the loss function or enforced in the network architecture \cite{raissi_physics_2019,fan_differentiable_2024}. 

All these models can be used in FSI-constrained structural optimization. However, simplified physical models can be inaccurate because the underlying assumptions are not satisfied in practical applications. For example, Narkhede {\it et al.} demonstrated that for a shock mitigation application, approximating the flow using empirical blast models can lead to $37\%$ error in the structure's maximum plastic strain \cite{narkhede_fluid_2025}. On the other hand, data-driven approaches typically require a large number of expensive CFD analyses to generate solution bases (for ROMs) and training data (for neural networks). The accuracy of these models depends heavily on how well the CFD analyses sample the relevant parameter space (e.g., operating conditions or structural geometry)~\cite{benner_survey_2015}. Because such models are constructed offline, prior to the optimization, the number of CFD analyses required for model generation can exceed that needed to complete the optimization directly using coupled CFD--CSD analyses. For example, in Bouhlel {\it et al}.~\cite{bouhlel2020scalable}, approximately $42,000$ CFD simulations were performed to train a deep learning surrogate for airfoil shape optimization. While these models become advantageous when reused across many optimization problems, achieving such reusability may demand an even broader set of pre-computed results.

In this paper, we present an adaptive, multi-fidelity optimization framework that integrates coupled CFD-CSD analysis with two additional components: 
\begin{enumerate}
	\item [(1)] a surrogate model of the fluid-induced loads on the structural surface, constructed and updated on-the-fly during the optimization process, and 
	\item [(2)] a decision model that estimates the surrogate's error and determines, for each design point, whether to employ the surrogate or the original CFD model.
\end{enumerate}
The optimization process begins with coupled CFD--CSD analyses to evaluate the initial design candidates. As the process advances, the resulting CFD data are progressively used to build and refine the surrogate model of the fluid loads. Because the available data grow only gradually with each iteration, the surrogate model should be lightweight and adaptive rather than heavily parameterized. Moreover, as the optimization converges, new query points are expected to cluster around local optima. Therefore, the surrogate model need not be uniformly accurate over the entire design space, but should provide sufficient accuracy in the regions actively explored by the optimizer. Based on these considerations, we employ nearest-neighbor search and unstructured local interpolation, which can be updated efficiently as new CFD data become available. As the optimization search becomes increasingly localized, CFD data are expected to accumulate locally, thereby improving the accuracy of  the surrogate model for subsequent design evaluations.

The decision model is constructed using Gaussian process regression (GPR). For each new design evaluation, it predicts whether the surrogate model is sufficiently accurate. If so, the surrogate is used to generate the fluid loads for the CSD analysis. Otherwise, the framework automatically reverts to the coupled CFD--CSD analysis, and its result is used to update both the surrogate and the decision model. GPR is well suited for this task because it provides both an estimate of the approximation error and a measure of the predictive uncertainty. In this work, the estimated error and associated uncertainty are used jointly to assess the accuracy of the surrogate model in a risk-aware manner. Moreover, the kernel function in GPR provides a flexible means to represent high-dimensional error fields as functions of the design variables, and its parameters can be calibrated progressively, allowing the decision model to be updated as the optimization proceeds.

Compared to relying completely on coupled CFD--CSD analyses, this adaptive, multi-fidelity framework is expected to maintain accuracy while reducing computational cost. It reuses the data accumulated from expensive simulations, which would otherwise be discarded after each iteration. Compared to SBAO and data-driven fluid dynamics models, this framework requires no pre-computation or offline training, preserves the high-fidelity CSD model, and continues to utilize the high-fidelity CFD model throughout the optimization process. The cost of updating and evaluating the surrogate and decision models is negligible compared to running a CFD analysis. Therefore, even in the worst case---when all design points are evaluated using coupled CFD--CSD analysis---the computational overhead remains trivial.


This framework is related to prior research on online model adaptation and adaptive fidelity selection in optimization (e.g., \cite{ong_evolutionary_2003,jeong_efficient_2005, gramacy_adaptive_2009,zahr2015progressive,peherstorfer2018survey,yong2019multi}). In particular, neural network-based approaches have been extensively explored for multi-fidelity modeling and data fusion, including convolutional neural networks, neural operators, transfer learning, and Bayesian neural networks \cite{conti2024multi,tao2024multi,xie2025multi,xue2026masked}. Another strategy is to learn the discrepancy between the surrogate and the high-fidelity model during the optimization process. The resulting error model is then applied as a bias correction to the surrogate's prediction, leaving the surrogate itself unchanged (e.g., \cite{lykkegaard2021accelerating,lykkegaard2023multilevel}).   These approaches provide alternative means for constructing surrogate and decision models. In comparison, the present framework is formulated specifically for FSI problems and exploits the disparity between fluid and structural computational costs to balance accuracy and computational efficiency. While the framework is conceptually applicable to both gradient-based and gradient-free optimization methods, this paper focuses on the latter, particularly genetic algorithms. Gradient-based FSI optimization is still uncommon due to the challenges of computing derivatives across coupled CFD--CSD solvers, although some recent studies have made progress in this area (e.g.,~\cite{sanchez2018coupled}). 

The remainder of this paper is organized as follows. In \sectionname~\ref{sec:optimization}, we formulate the constrained optimization problem for structures subject to FSI. We also use a simplified model to  justify the proposed approach by relating its performance to the added mass effect. In \sectionname~\ref{sec:framework}, we present the proposed optimization framework, focusing on the surrogate model of fluid-induced loads and the GPR-based decision model. Section~\ref{sec:implementation} describes the algorithms for implementing the framework and analyzes their computational complexity. Section~\ref{sec:results} presents two numerical experiments that demonstrate the performance of the framework. Finally, concluding remarks are provided in \sectionname~\ref{sec:conclusion}.

\section{Problem Description}
\label{sec:optimization}

\subsection{Optimization of structures subject to FSI}
\label{sec:definitions}
In general, PDE constraints can be handled in two ways. One approach is to treat them explicitly as equality constraints, known as simultaneous analysis and design (SAND). The other approach, referred to as nested analysis and design (NAND), is to treat the PDE as a function that maps the design variables of the optimization problem to the state variables of the PDE. Following the NAND approach, the optimization of a solid structure subject to FSI can be written as
	\begin{equation}
		\begin{aligned}
			\operatorname*{minimize}\limits_{\bm{\alpha}\in\Omega_\text{D}}			
			\quad
			& c\big(\boldsymbol{w}_\text{S}(\bm{\alpha}),~\boldsymbol{\alpha}\big),\\[2pt]
			\operatorname*{subject~to} \quad & \bm{g}\big(\boldsymbol{w}_\text{S}(\bm{\alpha}),~\boldsymbol{\alpha}) \le \bm{0}, \\[2pt]
			& \bm{h}\big(\boldsymbol{w}_\text{S}(\bm{\alpha}),~\boldsymbol{\alpha}\big) = \bm{0},
		\end{aligned}
		\label{eq:general_optimization_problem}
	\end{equation}
where $\bm{w}_\text{S}$ denotes the structural state variables. $\bm{\alpha}$ represents the design variables, such as the structure's geometric and material parameters. $c$ is the objective function being minimized. $\bm{g}$ and $\bm{h}$ denote the inequality and equality constraints, other than the governing PDEs. In this work, we assume that $c$, $\bm{g}$, and $\bm{h}$ do not depend explicitly on fluid variables.

\begin{figure}[H]
	\centering
	\includegraphics[width=0.75\textwidth]{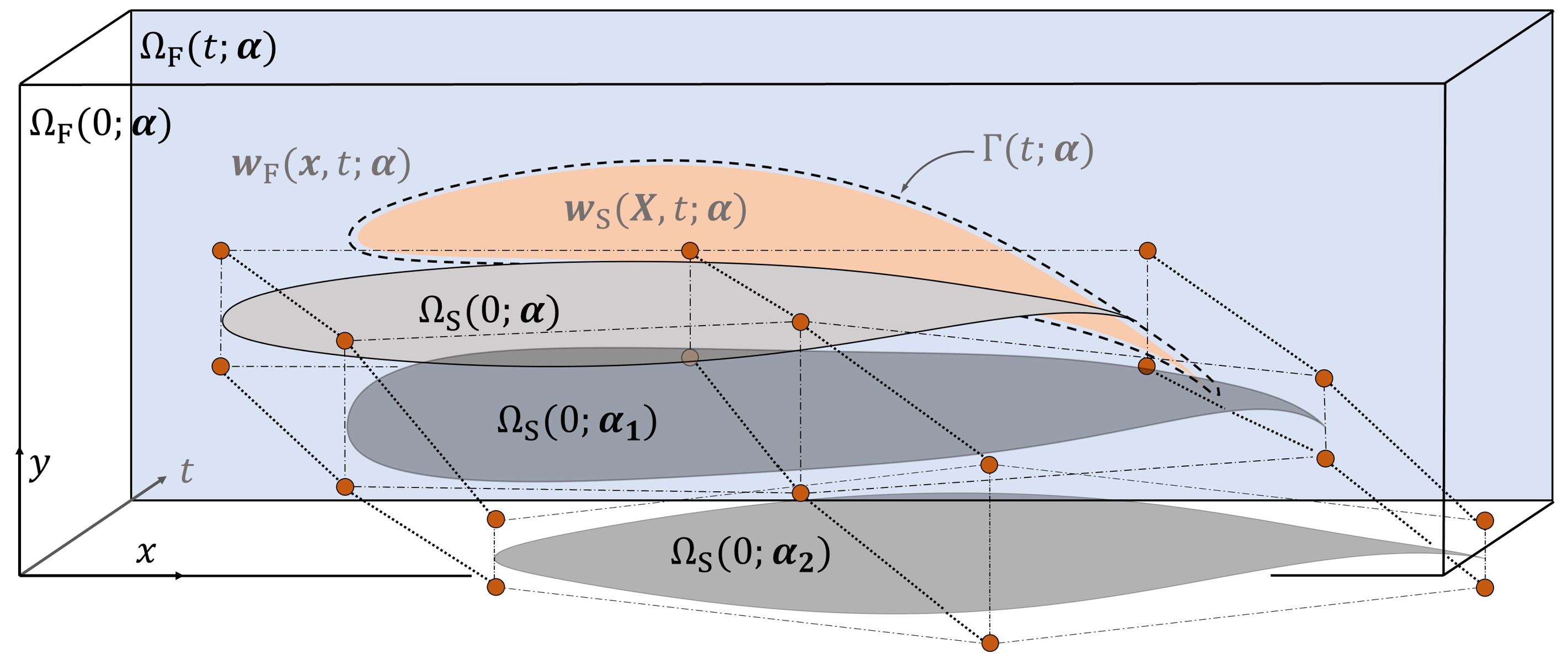}
	\caption{Fundamental concepts in FSI-constrained structural optimization: independent variables (space $\bm{x}$, time $t$, and design variables $\bm{\alpha}$), Eulerian and Lagrangian coordinates ($\bm{x}$ and $\bm{X}$), fluid and structural domains ($\Omega_{\mathrm{F}}$ and $\Omega_{\mathrm{S}}$), interface ($\Gamma$), and state variables ($\bm{w}_{\mathrm{F}}$ and $\bm{w}_{\mathrm{S}}$).}	
	\label{fig:illustration}
\end{figure}

Figure~\ref{fig:illustration} illustrates the problem setup. For a given design characterized by $\bm{\alpha}$, $\bm{w}_\text{S}$ is determined implicitly by the coupled FSI problem,
\begin{equation}
\mathcal{R}\big(\boldsymbol{w}_\text{F},~\boldsymbol{w}_\text{S};~\boldsymbol{\alpha}\big)=\bm{0},
\label{eq:coupled_pde_0}
\end{equation}
which includes the governing PDEs of fluid and structural dynamics and their initial and boundary conditions. Here, $\bm{w}_\text{F}$ denotes the fluid state variables, which are obtained simultaneously with $\bm{w}_\text{S}$. This coupled problem can be decomposed into a fluid sub-problem and a structural sub-problem, i.e.,

	\begin{equation}
		\mathcal{R}\big(\boldsymbol{w}_\text{F},~\boldsymbol{w}_\text{S};~\boldsymbol{\alpha}\big) :=	
		\begin{cases}
			\mathcal{R}_\text{F}\big(\boldsymbol{w}_\text{F},~\boldsymbol{w}_\text{S}\big)=\bm{0}, \\[2pt]
			
			\mathcal{R}_\text{S}\big(\boldsymbol{w}_\text{F},~\boldsymbol{w}_\text{S};~\boldsymbol{\alpha}\big)=\bm{0}. \\
		\end{cases}
		\label{eq:coupled_pde}
	\end{equation}

The fluid sub-problem can be written in its strong form as
\begin{equation}
		\mathcal{R}_\text{F}:\quad
		\begin{cases}		
			\dfrac{\partial \bm{w}_\text{F}}{\partial t} = \mathcal{L}(\bm{w}_\text{F}) + \bm{s}_\text{F}(\bm{w}_\text{F},\bm{x},t), & \forall \bm{x} \in \Omega_\text{F}(t), \\
           \mathcal{B}_\text{F}(\bm{w}_\text{F}) = \bm{0}, & \forall \bm{x} \in \partial\Omega_\text{F}(t) \setminus \Gamma(t), \\
			\bm{w}_\text{F}(\bm{x}, 0) = \bm{w}_{\text{F},0}(\bm{x}), & \forall \bm{x} \in \Omega_\text{F}(0), \\
			\phi_1 (\bm{w}_\text{F},~\bm{w}_\text{S})=\bm{0}, & \forall \bm{x}\in \Gamma(t),
		\end{cases}
		\label{eq:fluid_coupled_equation}
\end{equation}
where $\Omega_\text{F}(t)$ denotes the time-dependent fluid domain, and $\Gamma(t)$ the fluid-structure interface that evolves due to structural motion and deformation.  $\mathcal{L}$ is the spatial differential operator in the PDE, and $\bm{s}_\text{F}$ represents the source terms. For example, for a compressible inviscid flow governed by Euler equations,
	\begin{equation}		
		\bm{w}_\text{F} = \begin{bmatrix}
			\rho\\
			\rho\boldsymbol{v}\\
			E_t
		\end{bmatrix}, 
		\qquad
		\mathcal{L}\big(\bm{w}_\text{F}\big) =-\nabla\cdot
		\begin{bmatrix}
			\rho \boldsymbol{v}^T\\
			\rho \boldsymbol{v}\otimes \boldsymbol{v} + p\mathbb{I}\\
			\big(E_t + p\big)\boldsymbol{v}^T
		\end{bmatrix},
		\qquad
		\bm{s}_\text{F}(\bm{w}_\text{F},\bm{x},t)=\bm{0},
		\label{eq:euler_terms}
	\end{equation}
where $\rho$, $\bm{v}$, $p$, and $E_t$ represent fluid density, velocity, pressure, and total energy per unit volume, respectively. 

In~\eqref{eq:fluid_coupled_equation}, $\mathcal{B}_\text{F}$ represents the boundary conditions on the boundaries other than the fluid-structure interface ($\Gamma(t)$). $\bm{w}_{\text{F},0}$ is the initial condition, and $\phi_1$ represents the kinematic interface condition on $\Gamma(t)$. For example, for an inviscid flow,
\begin{equation}
\phi_1(\bm{w}_\text{F},\bm{w}_\text{S})= \Big(\bm{v} - \dfrac{\partial\bm{w}_\text{S}}{\partial t}\Big)\cdot\bm{n},
\label{eq:kinematic_interf_cond}
\end{equation}
where $\bm{w}_\text{S}$ denotes the structural displacement, and $\bm{n}$ the normal direction of the interface.


Unlike the fluid, the kinematics of solid structures is typically described in the Lagrangian reference frame. Let $\Omega_\text{S}(t)$ represent the structural configuration at time $t$. Based on the reference configuration, $\Omega_\text{S}(0)$, the structural sub-problem in \eqref{eq:coupled_pde} can be written as
\begin{equation}
		\mathcal{R}_\text{S}:\quad
		\begin{cases}
				\rho_\text{S}\big(\boldsymbol{X}\big)\dfrac{\partial^2\boldsymbol{w}_\text{S}}{\partial t^2}=\mathcal{D}_{\bm{X}}\big(J^{-1}\bm{F}\cdot\bm{S}\cdot\bm{F}^T\big) +\bm{s}_\text{S}\big(\bm{X},~t\big),
			& \forall \boldsymbol{X} \in \Omega_\text{S}(0), \\
			\mathcal{B}_\text{S}(\bm{w}_\text{S}) = \bm{0}, & \forall \bm{X}\in \partial\Omega_\text{S}(0)\setminus \Gamma(0) \\	
			\boldsymbol{w}_\text{S}\big(\boldsymbol{X},~ 0\big)=\boldsymbol{w}_{\text{S},0}\big(\boldsymbol{X}\big), & \forall \boldsymbol{X} \in \Omega_\text{S}(0), \\
			\dfrac{\partial\boldsymbol{w}_\text{S}}{\partial t}\big(\boldsymbol{X},~ 0\big)=\boldsymbol{v}_0\big(\boldsymbol{X}\big), & \forall \boldsymbol{X} \in \Omega_\text{S}(0),\\
			\phi_2(\bm{w}_\text{F},\bm{w}_\text{S}) = \bm{0}, & \forall \bm{X} \in \Gamma(0),
		\end{cases}
		\label{eq:solid_ibvp}
\end{equation}
where $\boldsymbol{X}$ denotes the material coordinates, $\rho_\text{S}$ is the structural density,  and $\boldsymbol{F}$ is the deformation gradient, with $J=\operatorname*{det}\big(\boldsymbol{F}\big)$. $\bm{S}$ represents the second Piola-Kirchhoff stress tensor, which is related to $\bm{w}_\text{S}$ through the constitutive material models. $\mathcal{D}_{\boldsymbol{X}}$ is the divergence operator expressed in Lagrangian coordinates. $\mathcal{B}_\text{S}$ represents the boundary conditions on the boundaries other than the fluid-structure interface. $\boldsymbol{w}_{\text{S},0}$ and $\boldsymbol{v}_0$ specify the initial displacement and velocity, respectively. The term $\boldsymbol{s}_\text{S}$ represents any external source terms acting on the structure.

$\phi_2$ denotes the dynamic interface condition at the fluid-structure interface. For example, for inviscid fluid flows
\begin{equation}
\phi_2(\bm{w}_\text{F},\bm{w}_\text{S}) = \big(\boldsymbol{\sigma}_\text{S} - p\mathbb{I}\big)\cdot \boldsymbol{n},
\label{eq:dynamic_interf_cond}
\end{equation}
where $\boldsymbol{\sigma}_\text{S}$ is the Cauchy stress tensor of the structure, i.e.,	
	\begin{equation}
		\boldsymbol{\sigma}_\text{S}=\dfrac{1}{J}\boldsymbol{F}\cdot\boldsymbol{S}\cdot\boldsymbol{F}^T.
	\end{equation}
	

\subsection{Overview of computational approach}

%

Although gradient-based methods are generally more efficient for continuous optimization problems, applying them to \eqref{eq:general_optimization_problem} is challenging. It requires computing $\mathrm{d}\bm{w}_\text{S}/\mathrm{d}\bm{\alpha}$ from the coupled system $\mathcal{R}$, which involves evaluating many derivatives and solving a coupled system across the fluid and structural sub-problems. Moreover, if $\mathcal{R}_\text{F}$ and $\mathcal{R}_\text{S}$ involve discontinuities, some derivatives may not exist.

In this work, we focus on gradient-free optimization methods. Algorithm~\ref{alg:gradient_free_fsi} outlines the general procedure. Starting from a set of $M$ design candidates, each optimization iteration consists of evaluating the candidates through FSI analyses, checking the stopping criterion, and, if necessary, generating a new set of candidates. Specifically, for each candidate design, the FSI problem is solved to obtain the corresponding fluid and structural states (Line 3), which are then used to evaluate the objective function and constraints (Line 4). After all $M$ candidates have been evaluated, the user-specified stopping criterion is checked. If it is satisfied, the final design is selected from the current candidates based on the penalized objective function (Line 7). Otherwise, a gradient-free update rule is applied to generate the design candidates for the next iteration (Line 9), and the process is repeated. The specific update rule depends on the optimization algorithm employed. For example, in a genetic algorithm, this step involves three operations known as selection, crossover, and mutation, which utilize the values of $c$, $\bm{g}$, and $\bm{h}$.

\begin{algorithm}[h]
\caption{Gradient-Free Optimization of Structures under Fluid-Structure Interaction}
\label{alg:gradient_free_fsi}
\begin{algorithmic}[1]
\Require Initial design candidates ${\bm{\alpha}^{(0)}_1, \dots, \bm{\alpha}^{(0)}_M}$, maximum optimization iteration $N_\text{iter}$, optimization stopping criterion (e.g., change in objective function's value)
\For{$k = 1, \dots, N_{\text{iter}}$}
\For{$i = 1, \dots, M$}
    \State Solve FSI problem $\mathcal{R}(\bm{w}_{\text{F}},~\bm{w}_{\text{S}};~\bm{\alpha}^{(k)}_i) = \bm{0}$ for $\bm{w}_{\text{F}}$ and $\bm{w}_{\text{S}}$.
    \State Evaluate objective function $c(\bm{w}_{\text{S}},~\bm{\alpha}^{(k)}_i)$ and constraints $\bm{g}(\bm{w}_{\text{S}},~\bm{\alpha}^{(k)}_i)$, $\bm{h}(\bm{w}_{\text{S}},~\bm{\alpha}^{(k)}_i)$.
\EndFor
    \State Check stopping criterion. If satisfied,\vspace{2mm}
    \State \quad\quad $\bm{\alpha}_\text{opt} = \underset{\left\{\bm{\alpha}_i^{(k)},~i=1,\dots,M\right\}}{\arg\min}~\left[ c(\bm{w}_\text{S}, \bm{\alpha}_i^{(k)}) + \mu \left( \norm{\max(\bm{0}, \bm{g}(\bm{w}_\text{S}, \bm{\alpha}_i^{(k)}))}^2 + \norm{\bm{h}(\bm{w}_\text{S}, \bm{\alpha}_i^{(k)})}^2 \right) \right]$. \Comment{$\mu$: penalty factor}\vspace{2mm}
    \State \quad\quad \textbf{break}
    \State Apply a gradient-free update rule to obtain new design candidates $\{\bm{\alpha}^{(k+1)}_1, \dots, \bm{\alpha}^{(k+1)}_M\}$.
\EndFor
\Ensure Final design: $\bm{\alpha}_\text{opt}$
\end{algorithmic}
\end{algorithm}

In most cases, the FSI problem $\mathcal{R}(\bm{w}_\text{F},\bm{w}_\text{S};~\bm{\alpha}_i^{(k)})=\bm{0}$ (Line 3 of Algorithm~\ref{alg:gradient_free_fsi}) can only be solved approximately using numerical methods. While the discretized fluid and structural governing equations can in principle be solved simultaneously using a monolithic approach \cite{hron2006monolithic,colomes2023monolithic}, partitioned procedures are more commonly adopted as they preserve the modularity of existing CFD and CSD solvers \cite{felippa2001partitioned,farhat_robust_2010}. A basic partitioned procedure can be described as follows. For time step $n=0,1,2,\dots$,
\begin{align}
\bm{W}_\text{F}^{n+1} &= \widetilde{\mathcal{R}}_\text{F}\big(\bm{W}_\text{F}^{n},~\bm{W}_\text{S}^{n}\big),\label{eq:discrete_coupled_problem_1}\\
\bm{W}_\text{S}^{n+1} &= \widetilde{\mathcal{R}}_\text{S}\big(\bm{W}_\text{F}^{n},~\bm{W}_\text{S}^{n};~\bm{\alpha}\big),
\label{eq:discrete_coupled_problem_2}
\end{align}
where $\bm{W}_\text{F}$ and $\bm{W}_\text{S}$ represent the  discretized fluid and structural state variables. At each time step, the CSD solver sends the displacement and velocity of its wetted surface (i.e., $\Gamma(t)$) to the CFD solver. These quantities define the fluid boundary position and boundary conditions for the current time step, thereby enforcing the kinematic interface condition ($\phi_1$ in~\eqref{eq:fluid_coupled_equation}). In return, the CFD solver sends the fluid-induced loads (i.e., distributed forces exerted by the fluid on the structure) to the CSD solver, which enforces the dynamic interface condition ($\phi_2$ in~\eqref{eq:solid_ibvp}). 

The above description represents a simplified partitioned procedure that assumes a one-step time integrator for both solvers and neglects coupling strategies such as predictor-corrector schemes, sub-iterations, and grid staggering. In the numerical experiments presented in this paper, the CFD solver M2C~\cite{zhao2026m2c} and the CSD solver Aero-S~\cite{aeros} are coupled through a staggered partitioned procedure developed in \cite{farhat_robust_2010} for compressible flow problems.  In this procedure, the fluid and structural time grids are offset by half a time step to improve accuracy and numerical stability.

\subsection{Motivation for current work}
\label{sec:motivation}

The solution of the coupled FSI problem dominates the overall cost of the optimization algorithm. A single CFD--CSD simulation can require on the order of $10^4$ CPU core-hours, if not more.  Algorithm~\ref{alg:gradient_free_fsi} may require thousands of such analyses. In comparison, the combined cost of all other steps in Algorithm~\ref{alg:gradient_free_fsi} is negligible. As mentioned in \sectionname~\ref{sec:introduction}, the cost of solving an FSI problem is typically dominated by the fluid sub-problem. In practice, it is not uncommon for the fluid solver to account for over $95\%$ of the total computational cost. Therefore, accelerating the fluid solution---while leaving the structural dynamics solver unchanged---has the potential to yield a substantial speedup in the overall optimization process.

Nonetheless, replacing the high-fidelity CFD solver with a surrogate model introduces errors in the fluid-induced loads. The impact of these errors is problem specific. In the following, we analyze a simple model problem to gain some insight.

Figure~\ref{fig:euler_beam_example} shows the geometry of the problem. The structure is an Euler--Bernoulli beam whose cross-sectional area and flexural rigidity are allowed to vary along its length, parameterized by the design variable $\bm{\alpha}$. The beam interacts with a fluid flow on one side. The corresponding fluid and structural sub-problems are defined as follows:

\begin{figure}[!htbp]
	\centering
	\includegraphics[width=0.3\textwidth]{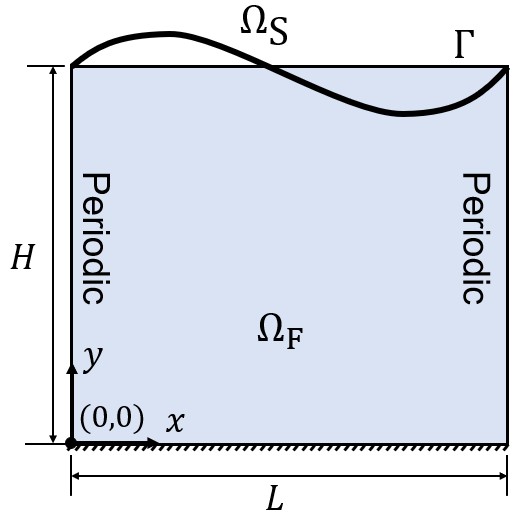}
	\caption{A two-dimensional fluid-structure interaction model problem.}
	\label{fig:euler_beam_example}
\end{figure}

\begin{equation}
	R_\text{F} : \begin{cases}
		&\nabla\cdot \bm{v} = 0,\\
		&\rho_\text{F}\dfrac{\partial\bm{v}}{\partial t} + \nabla p = 0,\\
		&p(0, y, t) = p(L, y, t),\\
		&\dfrac{\partial p(0,y,t)}{\partial x} = \dfrac{\partial p(L,y,t)}{\partial x}, \\
		&\bm{v}(x,0,t)\cdot\bm{n} = 0,\\
		&\bm{v}(x,H,t)\cdot\bm{n} = \dfrac{\partial w_\text{S}}{\partial t}.
	\end{cases}
\quad\quad
	R_\text{S} : \begin{cases}
		&\rho_\text{S}A(x;\boldsymbol{\alpha})\dfrac{\partial^2w}{\partial t^2} + \dfrac{\partial^2}{\partial x^2}\left(D(x; \boldsymbol{\alpha})\dfrac{\partial^2 w}{\partial x^2}\right) = f_\text{ext}, \\
		&w = 0, ~ \dfrac{\partial^2 w}{\partial x^2} = 0, ~ \text{for}~ x=0 ~\text{and}~ x = L,\\
		&w(x, 0) = w_0\sin(\pi x/L), \\
		&\dfrac{\partial w(x,0)}{\partial t} = 0,\\
		&f_\text{ext} = bp(x, H, t).
	\end{cases}
\end{equation}

To allow analytical solution, the fluid model has been simplified by assuming an incompressible, inviscid, and low-speed flow~\cite{Paidoussis2013FSI,cao2018robin}. While the kinematic interface condition on velocity is enforced, the interface position is assumed to be fixed at $y=H$. Combining the two fluid governing equations, we can eliminate velocity $\bm{v}$ to obtain a single Laplace equation for the pressure,
\begin{equation}
\nabla\cdot\nabla p = 0.\label{eq:fluid_laplace}
\end{equation}

The boundary conditions can also be expressed in $p$. In particular, at the fluid-structure interface ($y=H$), the kinematic interface condition becomes
\begin{equation}
\dfrac{\partial p}{\partial y} = -\rho_\text{F}\dfrac{\partial^2 w}{\partial t^2}.
\label{eq:converted_interface_condition}
\end{equation}

We seek separable solutions of the form
\begin{align}
	w(x,t) &= \hat{w}(t)\phi(x), \\
	p(x,y,t) &= \hat{p}(y,t) \phi(x),\label{eq:pressure_separation}
\end{align}
with $\phi(x) = \sin(\pi x/L)$ to satisfy the initial condition and the periodic boundary conditions. Substituting this ansatz into~\eqref{eq:fluid_laplace} and its boundary conditions at $y=0$ and $H$ yields 
\begin{equation}
	\hat{p}(H, t) = -\underbrace{\dfrac{\rho_\text{F} L}{\pi \tanh\left(\sfrac{\pi H}{L}\right)}}_{\displaystyle := M_a} \ddot{\hat{w}}(t),
	\label{eq:pressure_intermediate_sol}
\end{equation}
where $M_a$ represents the added mass effect of the fluid. Notably, $M_a$ depends only on the fluid density and geometry, and is independent of the structural design variable $\bm{\alpha}$. Integrating the structural governing equation over the spatial domain $0<x<L$ leads to
\begin{equation}
	\bar{M} \ddot{\hat{w}}(t) + \bar{D} \hat{w}(t) = \hat{p}(H, t),
	\label{eq:beam_ode}
\end{equation}
where the effective mass and stiffness are given by
\begin{equation}
	\bar{M}\big(\boldsymbol{\alpha}\big) = \dfrac{2}{bL}\int_{0}^{L}\rho_\text{S}\big(\phi(x)\big)^2A(x;\boldsymbol{\alpha}) dx,\quad\quad
	\bar{D}\big(\boldsymbol{\alpha}\big) = \dfrac{2}{bL}\int_{0}^{L}\big(\phi^{\prime\prime}(x)\big)^2D(x;\boldsymbol{\alpha}) dx.\label{eq:M_and_D}
\end{equation}

Substituting \eqref{eq:pressure_intermediate_sol} into \eqref{eq:beam_ode} yields the exact solution for the beam deflection,
\begin{equation}
w_\text{true} = w_0\sin\Big(\dfrac{\pi x}{L}\Big) \cos(\omega t),
\end{equation}
with
\begin{equation}
\omega = \sqrt{\dfrac{\bar{D}}{\bar{M}+M_a}}.
\end{equation}
\vspace{2mm}

We now consider the scenario in which the structural sub-problem $\mathcal{R}_\text{S}$ remains the same, while the fluid sub-problem $\mathcal{R}_\text{F}$ is replaced by a surrogate model constructed via localized approximation in the design space. Without loss of generality, we assume that the applied pressure corresponds to the exact solution of the FSI problem evaluated at a nearby design point, $\bm{\alpha}_1$, representing a constant extrapolation in the design space. In this case, the structural equation \eqref{eq:beam_ode} becomes
\begin{equation}
	\bar{M} \ddot{\hat{w}}(t) + \bar{D} \hat{w}(t) = w_0 M_a \omega_1^2 \cos(\omega_1  t),
	\label{eq:beam_ode_var}
\end{equation}
where $\omega_1 = \sqrt{\bar{D}_1/(\bar{M}_1+M_a)}$ is the beam's vibration frequency obtained with $\bm{\alpha}_1$. $\bar{M}_1$ and $\bar{D}_1$ have the same forms as in \eqref{eq:M_and_D}, with $\bm{\alpha}$ substituted by $\bm{\alpha}_1$. 


Solving \eqref{eq:beam_ode_var} yields its exact solution
\begin{equation}
w_\text{approx} = w_0 \sin\Big(\dfrac{\pi x}{L}\Big)\Bigg[\Big(1-\dfrac{M_a \omega_1^2}{\bar{D}-\omega_1^2 \bar{M}}\Big)\cos(\omega_\text{dry} t)
+ \dfrac{M_a \omega_1^2}{\bar{D}-\omega_1^2 \bar{M}}\cos(\omega_1 t)\Bigg].
\end{equation}
The subscript ``approx'' indicates that this result is obtained using an approximated (surrogate) fluid pressure. $\omega_\text{dry} = \sqrt{\bar{D}/\bar{M}}$ represents the beam's vibration frequency in vacuum. 

Thus, the impact of the fluid approximation is characterized by
\begin{equation}
e(\Delta\bm{\alpha},t)= \dfrac{w_\text{approx} - w_\text{true}}{w_\text{true}},
\end{equation}
where $\Delta\bm{\alpha} = \bm{\alpha}_1 - \bm{\alpha}$. Given the periodic nature of the solution, we examine the error at early times, as errors neither grow nor decay asymptotically. Applying Taylor series expansion with respect to both $\Delta \bm{\alpha}$ and $t$ yields
\begin{equation}
e(\Delta\bm{\alpha},t) = -\dfrac{1}{2} t^2 \dfrac{\gamma}{1+\gamma}\nabla_{\bm{\alpha}}\Bigg(\dfrac{\bar{D}}{\bar{M}}\Bigg)\cdot\Delta\bm{\alpha} + O(\|\Delta\bm{\alpha}\|^2,t^4),
\label{eq:model_rel_error}
\end{equation}
where $\gamma = M_a/\bar{M}$ denotes the ratio of fluid added mass to the structural mass.

Equation~\eqref{eq:model_rel_error} has several implications that justify the use of surrogate fluid models based on interpolation in the design space. As expected, the leading-order error increases monotonically with the magnitude of the fluid's added mass effect. When the ratio $\gamma$ is small---for example, when the fluid density $\rho_\text{F}$ is much smaller than the structural density $\rho_\text{S}$---the error tends to zero in the limit $\gamma \to 0$. Moreover, the factor $\gamma/(1+\gamma)$ is strictly concave and bounded by $1$. Therefore, even for problems that exhibit strong added mass effect, the error induced by the surrogate fluid model remains bounded and is proportional to $\Delta\bm{\alpha}$.

\section{An Adaptive Multi-Fidelity Optimization Framework}
\label{sec:framework}

We introduce an interpolation-based surrogate model to approximate the fluid-induced loads at the fluid-structure interface. This approximation is used alongside the original CFD model during the optimization process, while the structural sub-problem $\mathcal{R}_\text{S}$ is always solved using the high-fidelity CSD solver. A decision model is developed to estimate the approximation error. For each design point evaluation requested by the optimization algorithm, the decision model selects between a coupled CFD--CSD analysis and a CSD-only analysis with fluid loads computed using the surrogate model. Whenever a coupled analysis is performed, the fluid result is used to adaptively improve the accuracy of the surrogate model and to refine the decision model.

\subsection{Approximation of fluid-induced loads}\label{sec:pressure_approx}

During the optimization process, suppose that coupled CFD--CSD analyses have been performed for a set of structural designs. The corresponding fluid solutions are stored and treated as ground-truth data. To approximate the fluid pressure for evaluating a new design $\bm{\alpha}_*$, we construct a local, dimension-by-dimension interpolation surrogate of the form
\begin{equation}
	\tilde{p}(\bm{x}, t_n; \bm{\alpha}_*) =
	\sum_{k=1}^{N_b}
	w_k^{(\bm{\alpha})}
	\left[
	\sum_{i=0,1}
	w_{k,i}^{(t)}
	\left(
	\sum_{j}
	w_{k,i,j}^{(\bm{x})}
	\, p(\bm{x}_j, t_{n_k+i}; \bm{\alpha}_k)
	\right)
	\right],
	\label{eq:eulerian_pressure_approx}
\end{equation}
where $\bm{x}$ denotes the Eulerian coordinates of a point on the discretized fluid-structure interface, such as a node or Gauss point. $t_n$ is a time step of the CSD solver. The quantity $p(\bm{x}_j, t_{n_k+i}; \bm{\alpha}_k)$ represents the pressure obtained from the coupled analysis for design $\bm{\alpha}_k$, at spatial location (e.g., a node) $\bm{x}_j$ and time step $t_{n_k+i}$. The coefficients $w^{(\bm{\alpha})}$, $w^{(t)}$, and $w^{(\bm{x})}$ denote the interpolation weights in the design space, temporal domain, and spatial domain, respectively.

In the design space, all design variables are first normalized to allow consistent distance measures across different dimensions. For the new design $\bm{\alpha}_*$, we identify its $N_b$ nearest neighbors from the set of previously evaluated designs, denoted by $\bm{\alpha}_k$, $k = 1,\dots,N_b$. $N_b$ can be prescribed directly or determined by a cut-off distance. This construction connects naturally to the model problem in Sec.~\ref{sec:motivation}. In the limiting case of $N_b=1$, the approximation reduces to a constant extrapolation of the pressure field from the nearest neighbor, with $\bm{\alpha}_1$ playing the same role as the ``nearby design point'' $\bm{\alpha}_1$ in Sec.~\ref{sec:motivation}.

A common approach for interpolating unstructured data is radial basis function (RBF) interpolation, in which the weights are obtained as
\begin{equation}
	w_k^{(\bm{\alpha})}
	=
	\sum_{m=1}^{N_b}
	\phi\!\left(\lVert \bm{\alpha}_* - \bm{\alpha}_m \rVert\right)
	\left(\bm{\Phi}^{-1}\right)_{mk},
\end{equation}
where $\lVert \cdot \rVert$ denotes the Euclidean norm, $\phi(\cdot)$ is a radial basis function, and $\bm{\Phi} \in \mathbb{R}^{N_\text{b} \times N_\text{b}}$ is the interpolation matrix with entries $\left[\bm{\Phi}\right]_{ij} = \phi\!\left(\lVert \bm{\alpha}_i - \bm{\alpha}_j \rVert\right)$.

However, standard RBF interpolation does not enforce positivity or boundedness of the interpolation weights. In practice, we have observed that during the early iterations of the optimization process when the available data are sparsely distributed, the query design point $\bm{\alpha}_*$ often lies outside the convex hull of the sampled designs. In such cases, RBF interpolation can produce negative or high-magnitude weights, leading to large errors. 	

To avoid this issue, we instead adopt a normalized distance-based interpolation scheme:
\begin{equation}
	w_k^{(\bm{\alpha})}
	=
	\frac{
		\phi\!\left(\lVert \bm{\alpha}_* - \bm{\alpha}_k \rVert\right)
	}{
		\sum_{m=1}^{N_b}
		\phi\!\left(\lVert \bm{\alpha}_* - \bm{\alpha}_m \rVert\right)
	}.
	\label{eq:interp_weights}
\end{equation}

This formulation guarantees that all interpolation weights are non-negative and collectively sum to unity, providing a  convex combination of the neighbor values. When the query design $\bm{\alpha}_*$ lies outside the region spanned by the neighbors, the scheme naturally assigns the highest weight to the nearest neighbor. The choice of radial basis function $\phi$ can vary, with common options including multi-quadratic functions, inverse multi-quadratic functions, thin-plate splines, and Gaussian functions. In the numerical tests presented in this paper, we use the Gaussian functions, given by
\begin{equation}
	\phi\big(\norm{\bm{\alpha}_*-\bm{\alpha}_k}\big) = \exp\biggl(-\dfrac{\norm{\bm{\alpha}_*-\bm{\alpha}_k}^2}{2l^2}\biggr),
	\label{eq:gaussian_rbf}
\end{equation}
where $l$ is a scaling factor, set in this work to be slightly larger than the maximum distance between design points that have been evaluated with coupled analysis. 


The temporal interpolation is performed using a first-order linear scheme. In~\eqref{eq:eulerian_pressure_approx}, $n_k$ denotes the time-step index from the coupled analysis for design $\bm{\alpha}_k$ that satisfies $t_{n_k} \le t_n < t_{n_k+1}$. The weights $w_{k,0}^{(t)}$ and $w_{k,1}^{(t)}$ are given by
\begin{equation}
	w_{k,i}^{(t)} =
	\frac{\big|\,t_n - t_{n_k+1-i}\,\big|}{t_{n_k+1} - t_{n_k}},
	\qquad i = 0,1.
\end{equation}

The spatial interpolation needs to properly account for the mismatches between different coordinate systems. In a coupled analysis, fluid variables such as pressure are computed at nodes (or cell centers) of an Eulerian mesh, which generally do not coincide with the nodes or Gauss points of the Lagrangian structural mesh. Moreover, the structural geometry may vary across different designs (e.g., $\bm{\alpha}_*$ and $\bm{\alpha}_k$), leading to discrepancies in the Lagrangian coordinates of the interface.

\begin{figure}[H]
	\centering
	\includegraphics[width=0.6\linewidth]{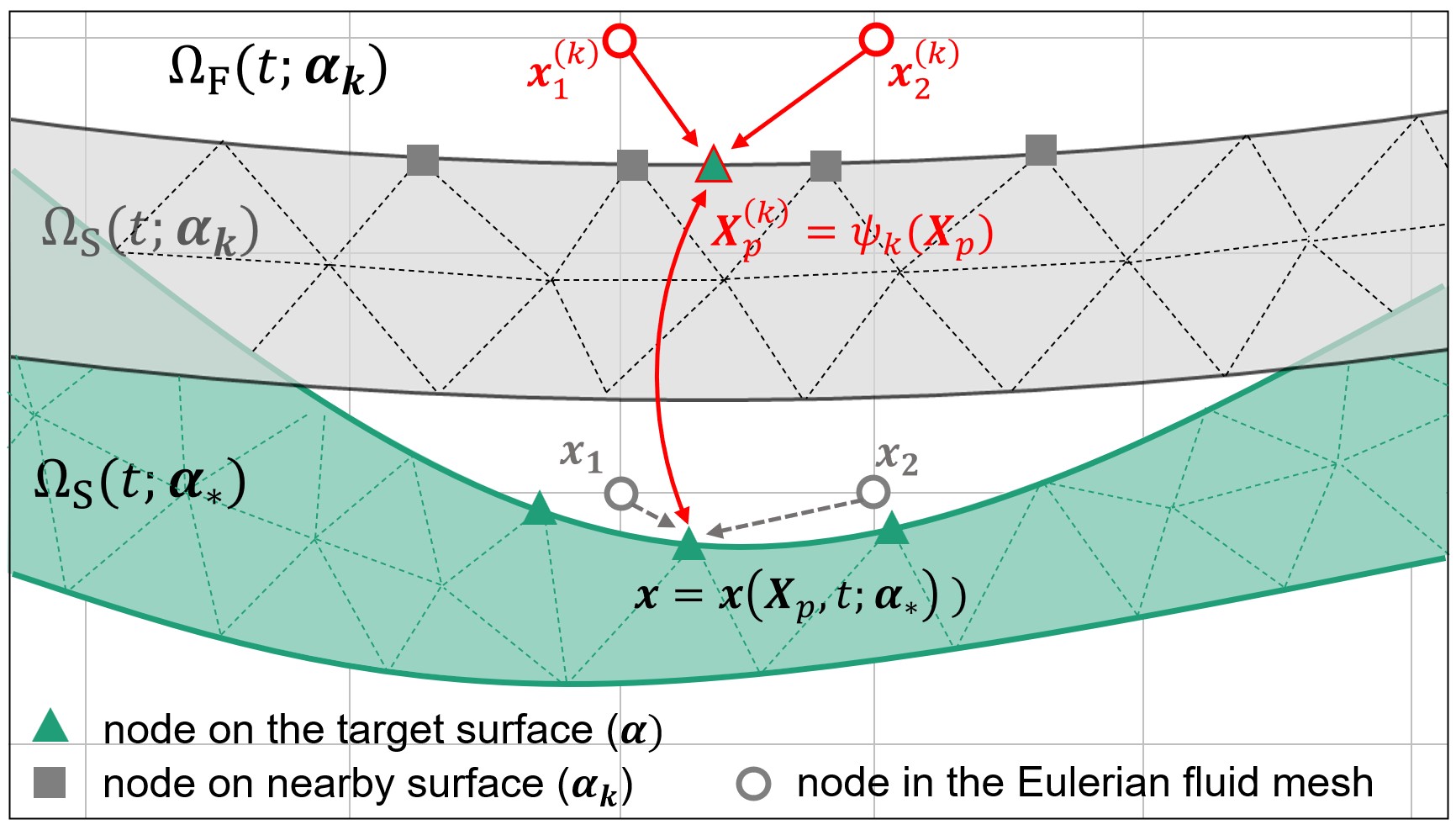}
	\caption{Illustration of spatial interpolation across different structural designs and between Eulerian fluid and Lagrangian structural reference frames.}
	\label{fig:mesh_neighbor_interpolation}
\end{figure}

A straightforward approach is to perform the interpolation entirely in the Eulerian reference frame by directly applying~\eqref{eq:eulerian_pressure_approx}, with $\bm{x}_j$ being fluid nodes from the coupled analysis for design $\bm{\alpha}_k$ that lie in the vicinity of the interface point $\bm{x}$. This approach has a clear drawback, as illustrated in \figurename~\ref{fig:mesh_neighbor_interpolation}. The figure shows the structural mesh of the query design $\bm{\alpha}_*$ overlaid with both the fluid and structural meshes corresponding to a nearby design $\bm{\alpha}_k$ used in the interpolation. Computing the pressure at $\bm{x}$ involves interpolating values from nearby nodes on the fluid mesh, such as those labeled $\bm{x}_{1}$ and $\bm{x}_{2}$. In the coupled analysis for design $\bm{\alpha}_k$, these nodes lie beneath the structure, near its lower surface, whereas in the query design the point $\bm{x}$ is located on the upper surface.  In other words, this would result in using fluid pressure on the wrong side of the structure.


In this work, spatial interpolation is performed using both Lagrangian and Eulerian coordinates. For each neighboring design  $\bm{\alpha}_k$, we define a function 
\begin{equation}
	\bm{X}^{(k)} = \psi_k(\bm{X})
	\label{eq:Lagrangian_mapping}
\end{equation}
that maps $\Gamma(0;\bm{\alpha}_*)$ to $\Gamma(0;\bm{\alpha}_k)$ using Lagrangian coordinates. $\psi_k$ is expected to be continuous in $\bm{X}$.  Furthermore, when all such functions are considered collectively as a function of the design variables, i.e., $\psi_k(\bm{X}) = \Psi(\bm{\alpha}_k,\bm{X})$, its dependence on $\bm{\alpha}$ is also expected to be continuous. As long as the design space does not involve topological changes to the interface geometry (i.e., the interfaces associated with all designs are homeomorphic), such functions exist and are usually straightforward to construct. For example, if the interface geometry remains unchanged across designs, $\psi_k$ reduces to the identity map. For designs involving shape changes—e.g., evolving from a sphere to an oval—the shape transformation itself defines $\Psi$. As illustrated in Fig.~\ref{fig:mesh_neighbor_interpolation}, the continuity requirements ensure that $\psi_k$ maps a point $\bm{X}_p$ on the lower surface of the query design $\bm{\alpha}^*$ to a point on the lower surface of $\bm{\alpha}_k$.

Once $\bm{X}^{(k)}$ is identified, its corresponding position, $\bm{x}(\bm{X}^{(k)}, t_{n_k+i}; \bm{\alpha}_k)$, is readily available from the structural displacement field provided by the coupled analysis for design $\bm{\alpha}_k$. The pressure at this location is then evaluated by interpolating from nearby nodes on the fluid mesh, such as those labeled $\bm{x}^{(k)}_1$ and $\bm{x}^{(k)}_2$. This interpolation can be performed using either the fluid mesh, or through an unstructured, distance-based approach similar to that used in the design space. The chosen strategy defines the spatial interpolation weights $w^{(\bm{x})}$.

Incorporating the mapping $\psi_k$, the pressure approximation in~\eqref{eq:eulerian_pressure_approx} is rewritten as
\begin{equation}
	\tilde{p}(\bm{X}, t_n; \bm{\alpha}_*) =
	\sum_{k=1}^{N_b}
	w_k^{(\bm{\alpha})}
	\left(
	\sum_{i=0,1}
	w_{k,i}^{(t)}
	\, \tilde{p}\!\left(\psi_k(\bm{X}), t_{n_k+i}; \bm{\alpha}_k\right)
	\right),
	\label{eq:pressure_formula}
\end{equation}
with
\begin{equation}
	\tilde{p}\!\left(\psi_k(\bm{X}), t_{n_k+i}; \bm{\alpha}_k\right)
	=
	\sum_{j=1}^{N_p}
	w_{k,i,j}^{(\bm{x})}
	\, p(\bm{x}_j, t_{n_k+i}; \bm{\alpha}_k).
	\label{eq:pressure_formula_helper}
\end{equation}
Here, $\bm{x}_j$, $j=1,\ldots,N_p$, denote the fluid mesh nodes used to interpolate pressure at $\bm{x}(\bm{X}^{(k)}, t_{n_k+i}; \bm{\alpha}_k)$. Once $\tilde{p}$ is computed, the distributed loads are obtained by integrating $\tilde{p}$ on the structural surface with finite element basis functions.

\vspace{2mm}

This method can be extended to fluid flows with viscous shear forces by approximating the surface traction vector $\bm{T}$ instead of pressure $p$. However, simply replacing $p$ by $\bm{T}$ in \eqref{eq:pressure_formula} and \eqref{eq:pressure_formula_helper} is not sufficient. Different structural designs generally have different surface orientations and normal directions at corresponding material points. Because the interpolation weights $w^{(\bm{\alpha})}_k$ are obtained solely from distances in the design space, they do not account for these geometric discrepancies. For example, consider a problem where the structure experiences purely normal traction (i.e., no shear). Suppose the query design $\bm{\alpha}_*$ has two neighbors, $\bm{\alpha}_1$ and $\bm{\alpha}_2$, with surface normals $\bm{n}_1$ and $\bm{n_2}$ that differ from the query design's surface normal $\bm{n}_*$. Direct interpolation in a global coordinate frame yields
\begin{equation}
	\tilde{\bm{T}}(\bm{X}, t; \bm{\alpha}_*) = w_1^{(\bm{\alpha})}
	\left[
	\sum_{i=0,1}
	w_{k,i}^{(t)}
	\, \left(
	\sum_{j}
	w_{k,i,j}^{(\bm{x})}
	\, p(\bm{x}_j, t_{n_k+i}; \bm{\alpha}_1)
	\right)
	\right]\bm{n}_1 + 
	w_2^{(\bm{\alpha})}
	\left[
	\sum_{i=0,1}
	w_{k,i}^{(t)}
	\, \left(
	\sum_{j}
	w_{k,i,j}^{(\bm{x})}
	\, p(\bm{x}_j, t_{n_k+i}; \bm{\alpha}_2)
	\right)
	\right]\bm{n}_2,
\end{equation}
which, in general, is not aligned with $\bm{n}_*$, and therefore introducing a spurious tangential component.

\begin{figure}[H]
	\centering
	\includegraphics[width=0.5\linewidth]{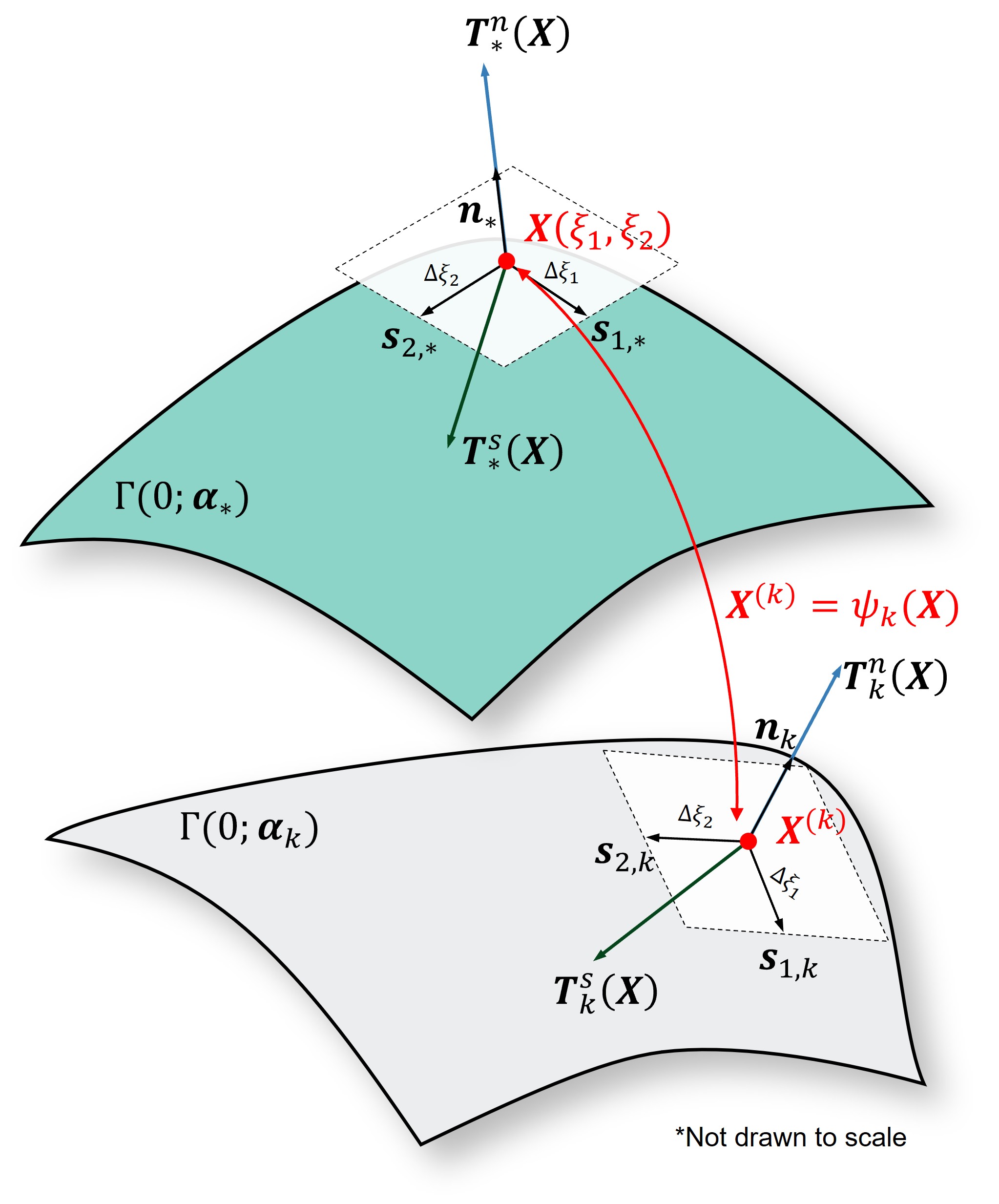}
	\caption{Illustration of the local orthonormal basis used for traction decomposition across structural designs. The normal traction component $\bm{T}^n(\bm{X}, t, \bm{\alpha}) = T^n\bm{n}$ acts perpendicular to the surface, while the shear traction component $\bm{T}^n(\bm{X}, t, \bm{\alpha}) = T^n\bm{n}$ and $\bm{T}^s(\bm{X}, t, \bm{\alpha}) = T^{s_1}\bm{s}_1 + T^{s_2}\bm{s}_2$ lies in the local tangent plane. Collectively, $\bm{T} = \bm{T}^n + \bm{T}^s$.}
	\label{fig:traction_components}
\end{figure}

To avoid this issue, we decompose the traction vector at each neighboring design into scalar components relative to a local orthonormal basis. These components physically represent the normal and tangential stresses in the local basis. Each scalar component is then interpolated separately, and the traction is then reconstructed using the local basis of the query design (\figurename~\ref{fig:traction_components}), effectively making the interpolation independent of geometric differences between structural designs.

The local orthonormal basis comprises the surface normal $\bm{n}$ and two tangent vectors, $\bm{s}_1$ and $\bm{s}_2$, defined on the tangent plane of the fluid-structure interface. While the surface normal is directly available from the geometry, the tangent vectors must be determined. Let $(\xi_1, \,\xi_2)$ be parametric coordinates on the reference interface $\Gamma(0; \alpha_*)$, with $\bm{X} = \bm{X}(\xi_1, \xi_2)$. We define the neighboring material points in each parametric direction as
\begin{equation}
	\bm{X}^{+1} = \bm{X}(\xi_1 + \Delta\xi_1, \xi_2), \quad \bm{X}^{+2} = \bm{X}(\xi_1, \xi_2 + \Delta\xi_2).
\end{equation}

For a neighboring design $\bm{\alpha}_k$, the two tangent vectors at the mapped point $\psi_k(\bm{X})$ are defined as
\begin{equation}
	\bm{s}_i\big(\psi_k(\bm{X}),\, t;\, \bm{\alpha}_k\big)
	= \frac{\bm{x}\big(\psi_k(\bm{X}^{+i}),\, t;\, \bm{\alpha}_k\big) - \bm{x}\big(\psi_k(\bm{X}),\, t;\, \bm{\alpha}_k\big)}
	{\left\|\bm{x}\big(\psi_k(\bm{X}^{+i}),\, t;\, \bm{\alpha}_k\big) - \bm{x}\big(\psi_k(\bm{X}),\, t;\, \bm{\alpha}_k\big)\right\|},
	\quad i = 1, 2,
\end{equation}
where $\bm{x}(\bm{X}, t; \bm{\alpha})$ denotes the spatial position of material point $\bm{X}$ at time $t$ for design $\bm{\alpha}$. Similarly, the tangent vectors for the query design are
\begin{equation}
	\bm{s}_i\big(\bm{X},\, t;\, \bm{\alpha}_*\big)
	= \frac{\bm{x}\big(\bm{X}^{+i},\, t;\, \bm{\alpha}_*\big) - \bm{x}\big(\bm{X},\, t;\, \bm{\alpha}_*\big)}
	{\left\|\bm{x}\big(\bm{X}^{+i},\, t;\, \bm{\alpha}_*\big) - \bm{x}\big(\bm{X},\, t;\, \bm{\alpha}_*\big)\right\|},
	\quad i = 1, 2.
\end{equation}

Using this local basis, the traction vector at each neighbor is decomposed as
\begin{equation}
	\bm{T}\big(\psi_k(\bm{X}), t;\bm{\alpha}_k\big)
	= T^n_k \bm{n}_k + T^{s_1}_k \bm{s}_{1,k} + T^{s_2}_k \bm{s}_{2,k},
\end{equation}
where $T^n_k$, $T^{s_1}_k$, and $T^{s_2}_k$ are the scalar components of the traction in the normal and tangential directions, respectively, with $\bm{n}_k = \bm{n}(\psi_k(\bm{X}), t; \bm{\alpha}_k)$ and $\bm{s}_{i, k} = \bm{s}(\psi_k(\bm{X}), t; \bm{\alpha}_k)$.

The traction components for a new structural design $\bm{\alpha}_*$ are approximated by interpolating each scalar component separately using the normalized distance-based interpolation, i.e.,
\begin{equation}
	\tilde{T}^{\chi}(\bm{X}, t_n; \bm{\alpha}_*) =
	\sum_{k=1}^{N_b}
	w_k^{(\bm{\alpha})}
	\left(
	\sum_{i=0,1}
	w_{k,i}^{(t)}
	\, \tilde{T}^{\chi}\!\left(\psi_k(\bm{X}), t_{n_k+i}; \bm{\alpha}_k\right)
	\right),\quad \chi=n,~s_1,~s_2,
	\label{eq:traction_formula}
\end{equation}
with
\begin{equation}
	\tilde{T}^{\chi}\!\left(\psi_k(\bm{X}), t_{n_k+i}; \bm{\alpha}_k\right)
	=
	\sum_{j=1}^{N_p}
	w_{k,i,j}^{(\bm{x})}
	\, T^{\chi}(\bm{x}_j, t_{n_k+i}; \bm{\alpha}_k).
	\label{eq:traction_formula_helper}
\end{equation}

The approximate traction vector is then reconstructed using the local basis of the query design:
\begin{equation}
	\tilde{\bm{T}}\big(\bm{X},\, t;\, \bm{\alpha}_*\big)
	= \tilde{T}^n \bm{n}_* + \tilde{T}^{s_1} \bm{s}_{1,*} + \tilde{T}^{s_2} \bm{s}_{2,*},
\end{equation}
where $\bm{n}_* = \bm{n}(\bm{X}, t; \bm{\alpha}_*)$ and $\bm{s}_{i,*} = \bm{s}_i(\bm{X}, t; \bm{\alpha}_*)$.

For fluid flows that only exert a pressure on the fluid-structure interface, $T^n = p$ and $T^{s_1} = T^{s_2} = 0$. Substituting these in \eqref{eq:traction_formula} and \eqref{eq:traction_formula_helper} recovers the pressure approximation given by \eqref{eq:pressure_formula}.

\subsection{Adaptive decision model}\label{sec:adaptive_model}
                                                                        
We develop an adaptive decision model based on Gaussian process regression (GPR) to estimate the accuracy of the surrogate model introduced in the previous subsection. Assuming the fluid-induced loads are computed using pressure only (i.e.,~\eqref{eq:pressure_formula}), we define the approximation error as
\begin{equation}
	e\big(\bm{\alpha}\big) = \dfrac{1}{N_t} \sum_{n=1}^{N_t}
	\dfrac{\norm{p\big(\bm{X},~t^{n};~\bm{\alpha}\big) -\tilde{p}\big(\bm{X},~t^{n};~\bm{\alpha}\big)}}{\norm{p\big(\bm{X},~t^{n};~\bm{\alpha}\big)}},
	\label{eq:rms_error}
\end{equation}
where $p$ and $\tilde{p}$ denote the fluid pressure field generated by coupled CFD--CSD analysis and the surrogate model, respectively. $\norm{\cdot}$ denotes the function $L^2$ norm over the spatial domain of $\bm{X}$. $N_t$ is the total number of time steps. Equation~\eqref{eq:rms_error} can be extended to the traction approximation in \eqref{eq:traction_formula} by generalizing the norm to vector-valued functions.

Let $\mathcal{A}=\left\{\bm{\alpha}_1, \bm{\alpha}_2, \dots, \bm{\alpha}_{N_\text{true}}\right\}$ denote the set of structural designs that have been evaluated using coupled CFD-CSD analyses. For each design $\bm{\alpha}_k\in\mathcal{A}$, we compute the approximate pressure field $\tilde{p}$ using the surrogate model and evaluate its error, $e_k = e(\bm{\alpha}_k)$, by \eqref{eq:rms_error}. These values are denoted collectively by vector $\bm{e}=\left[e_1, e_2, \dots,e_{N_\text{true}}\right]^T$. The regression dataset is then given by
\begin{equation}
	\mathcal{D} = \left\{(\bm{\alpha}_1,e_1), (\bm{\alpha}_2,e_2), \dots, (\bm{\alpha}_{N_\text{true}},e_{N_\text{true}})\right\}.
	\label{eq:regression_dataset}
\end{equation}

To predict the error at a new design point $\bm{\alpha}_*$, we employ the regression model
\begin{equation}
	e\big(\bm{\alpha}\big) = \bm{y}\big(\bm{\alpha}\big)^T\bm{\beta} + f\big(\bm{\alpha}\big).
\end{equation}

The first term, $\bm{y}(\bm{\alpha})^T\bm{\beta}$, captures the global trend in the dataset $\mathcal{D}$ while the second term, $f(\bm{\alpha})$, models the local deviations from this trend. Here, $\bm{y}(\bm{\alpha})$ is chosen to be a set of low-order polynomial basis functions and $\bm{\beta}$ their corresponding coefficients. For example, if $\bm{\alpha} = [\alpha_{(1)},\alpha_{(2)}]^T \in \mathbb{R}^2$, a second-order polynomial basis is given by $\displaystyle\bm{y} = \left[1, \alpha_{(1)}, \alpha_{(2)}, \alpha_{(1)}^2, \alpha_{(2)}^2, \alpha_{(1)}\alpha_{(2)}\right]^T$. The local deviation term is modeled using a Gaussian process (GP) with zero mean and a covariance function $k(\bm{\alpha}, \bm{\alpha}^\prime)$, i.e.,	
\begin{equation}
	f\big(\bm{\alpha}\big) \sim GP\big(0,~k\big(\bm{\alpha}, \bm{\alpha}^\prime\big)\big).
\end{equation}

By definition, a Gaussian process implies that any finite collection of function evaluations (here, $f$) follows a joint Gaussian distribution \cite{williams2006gaussian}. Consequently, the training error vector $\bm{e}$ and the error at the test point $e_* = e(\bm{\alpha}_*)$ satisfy
\begin{equation}
	\begin{bmatrix}
		\bm{e} \\
		e_*
	\end{bmatrix} \Bigg| ~ \mathcal{A},\bm{\beta}  \sim \mathcal{N}\left(
	\begin{bmatrix}
		\bm{Y}^T\bm{\beta} \\
		\bm{y}_*^T\bm{\beta}
	\end{bmatrix},~
	\begin{bmatrix}
		\bm{K} &\bm{k}_*\\
		\bm{k}_*^T &k_{**}
	\end{bmatrix}
	\right),
	\label{eq:joint_distrib}
\end{equation}
where  $\bm{Y} = \left[\bm{y}(\bm{\alpha}_1), \bm{y}(\bm{\alpha}_2), \dots, \bm{y}(\bm{\alpha}_{N_\text{true}})\right]$ collects the basis vectors $\bm{y}$ evaluated at training points, and $\bm{y}_* = \bm{y}(\bm{\alpha}_*)$. The covariance matrix $\bm{K} \in \mathbb{R}^{N_\text{true} \times N_\text{true}}$ has entries
\begin{equation}
\left[\bm{K}\right]_{ij} = k(\bm{\alpha}_i,\bm{\alpha}_j) + \sigma_m^2\delta_{ij},
\label{eq:K_entries}
\end{equation}
where $\sigma_m^2$ is the measurement noise variance, and $\delta_{ij}$ is the Kronecker delta. $\bm{k}_*$ is the cross-covariance vector, given by $\bm{k}_* = \left[k(\bm{\alpha}_*,\bm{\alpha}_1), k(\bm{\alpha}_*,\bm{\alpha}_2), \dots, k(\bm{\alpha}_*,\bm{\alpha}_{N_\text{true}})\right]^T$.  $k_{**} = k(\bm{\alpha}_*, \bm{\alpha}_*)$ is the prior variance at $\bm{\alpha}_*$.

Conditioning the joint distribution in \eqref{eq:joint_distrib} yields the following posterior distribution for $e_*$:
\begin{equation}
	e_* ~ \big| ~ \mathcal{A}, \bm{e}, \bm{\beta} \sim \mathcal{N}\left(\bm{y}_*^T\bm{\beta} + \bm{k}_*^T\bm{K}^{-1}(\bm{e} - \bm{Y}^T\bm{\beta}), ~k_{**} - \bm{k}_*^T\bm{K}^{-1}\bm{k}_*\right).
	\label{eq:posterior}
\end{equation}

By this distribution, the error $e_*$ depends on the choice of the covariance function $k(\cdot,\cdot)$, the measurement variance $\sigma_m^2$, and the polynomial coefficients $\bm{\beta}$. We treat $\bm{\beta}$ as an uncertain parameter and assign it a Gaussian prior, namely $\bm{\beta} \sim \mathcal{N}(\bm{b},~\bm{B})$. This treatment allows the model to account for parameter uncertainty, which is particularly important when the training dataset $\mathcal{D}$ is small during the early iterations of the optimization. By Bayes' theorem, the posterior distribution of $\bm{\beta}$ is given by
\begin{equation}
	\bm{\beta} ~ \big| ~ \mathcal{A}, \bm{e} \sim \mathcal{N}\left((\bm{B}^{-1}+\bm{Y}\bm{K}^{-1}\bm{Y}^T)^{-1}(\bm{Y}\bm{K}^{-1}\bm{
		e}+\bm{B}^{-1}\bm{b}), %
	~(\bm{B}^{-1}+\bm{Y}\bm{K}^{-1}\bm{Y}^T)^{-1}\right).
	\label{eq:beta_posterior}
\end{equation}

In this work, no prior information is available regarding the trend of the approximation error before the optimization begins. To reflect this in the model formulation, we adopt a vague prior by taking $\bm{B}^{-1} \rightarrow \bm{0}$. In this limit, the prior distribution  has no influence on the posterior inference of $\bm{\beta}$, so that their posterior distribution is completely determined by the dataset $\mathcal{D}$. This data-driven treatment is further justified in the present context because the approximation error defined in \eqref{eq:rms_error} depends on problem-specific factors---such as the sensitivity of the fluid loads to structural design variables---that is unknown \textit{a priori}. Under the non-informative prior, the posterior mean of $\bm{\beta}$ reduces to 
\begin{equation}
	\bar{\bm{\beta}} \equiv \mathbb{E}\big[\bm{\beta}~\big|~\mathcal{A},\bm{e}\big] = \big(\bm{Y}\bm{K}^{-1}\bm{Y}^T\big)^{-1}\big(\bm{Y}\bm{K}^{-1}\bm{e}\big).
	\label{eq:beta_bar}
\end{equation}

We marginalize the predictive distribution over the posterior distribution of $\bm{\beta}$. By \eqref{eq:posterior}, \eqref{eq:beta_posterior}, and \eqref{eq:beta_bar}, we obtain a Gaussian distribution for $e_*$ conditional on only the data set $\mathcal{D}$, with mean and variance given by
\begin{subequations}
	\begin{align}
		\mathbb{E}\big[{e}_*~\big|~\mathcal{A},\bm{e}\big] &= \mathbb{E}_{\bm{\beta}}\big[\mathbb{E}\big[e_*~\big|~\mathcal{A},\bm{e},\bm{\beta}\big]\big] = \bm{y}_*^T \bar{\bm{\beta}} + \bm{k}_*^T \bm{K}^{-1} \big(\bm{e} - \bm{Y}^T \bar{\bm{\beta}}\big), 
		\label{eq:gp_expectation} \\
		\mathbb{V}\big[{e}_*~\big|~\mathcal{A},\bm{e}\big] &= \mathbb{E}_{\bm{\beta}}\big[\mathbb{V}\big[e_*~\big|~\mathcal{A},\bm{e},\bm{\beta}\big]\big] + \mathbb{V}_{\bm{\beta}}\big[\mathbb{E}\big[e_*~\big|~\mathcal{A},\bm{e},\bm{\beta}\big]\big]\nonumber\\
		&=  k_{**} - \bm{k}_*^T \bm{K}^{-1} \bm{k}_* + \big(\bm{y}_* - \bm{Y} \bm{K}^{-1} \bm{k}_* \big)^T \big(\bm{Y} \bm{K}^{-1} \bm{Y}^T \big)^{-1} \big(\bm{y}_* - \bm{Y} \bm{K}^{-1} \bm{k}_* \big).
		\label{eq:gp_variance}
	\end{align}
\end{subequations}

This marginalization  allows  the resulting predictive variance to account for not only the intrinsic uncertainty of the Gaussian process but also the uncertainty in $\bm{\beta}$. $\mathbb{E}\big[{e}_*~\big|~\mathcal{A},\bm{e}\big]$ provides an estimate of the surrogate model's error for structural design $\bm{\alpha}_*$, while the corresponding uncertainty is quantified by the standard deviation
\begin{equation}
	u_* = \sqrt{\mathbb{V}\big[{e}_*~\big|~\mathcal{A},\bm{e}\big]}.
	\label{eq:uncertainty}
\end{equation}

The covariance function is chosen to be the squared exponential function,
\begin{equation}
	k\big(\bm{\alpha}, \bm{\alpha}^\prime\big) = \sigma^2\exp\bigg(-\dfrac{1}{2}\big(\bm{\alpha}-\bm{\alpha}^\prime\big)^T\bm{\Theta}\big(\bm{\alpha}-\bm{\alpha}^\prime\big)\bigg),
\end{equation}
where $\bm{\Theta}=\text{diag}\big(\theta_1,\cdots,\theta_d\big)$ is a diagonal matrix containing the inverse length-scale parameters associated with each design variable dimension, and $\sigma^2$ is the kernel variance. 

The parameters $\sigma_m^2$, $\{\theta_i\}_{i=1}^d$, and $\sigma^2$ are determined by maximizing the logarithmic marginal likelihood of the observed dataset $\mathcal{D}$. Under the assumption of a non-informative prior for $\bm{\beta}$,
\begin{equation}
p(\bm{e}~|~\mathcal{A})
=
\dfrac{1}{\sqrt{(2\pi)^{N_{\text{true}}-N_{\bm{\beta}}}|\bm{K}| |\bm{Y}\bm{K}^{-1}\bm{Y}^T|}}
\exp\left( 
-\dfrac{1}{2}(\bm{e}-\bm{Y}^T\bm{\bar{\beta}})^T\bm{K}^{-1}(\bm{e}-\bm{Y}^T \bm{\bar{\beta}})
\right),
\end{equation}
where $N_\beta$ denotes the dimension of $\bm{\beta}$, that is, the number of polynomial basis functions in $\bm{y}$. Maximizing the logarithmic likelihood yields 
\begin{equation}
	\left\{\sigma_m^2, \{\theta_i\}_{i=1}^d, \sigma^2\right\} = \arg\max~\left(-\dfrac{1}{2}(\bm{e}-\bm{Y}^T\bar{\bm{\beta}})^T\bm{K}^{-1}(\bm{e}-\bm{Y}^T\bar{\bm{\beta}}) - \dfrac{1}{2}\log\lvert\bm{K}\rvert - \dfrac{1}{2}\log\lvert\bm{Y}\bm{K}^{-1}\bm{Y}^T\rvert\right).
	\label{eq:hyperparameters}
\end{equation}
This unconstrained optimization problem can be solved numerically using gradient-based methods.

\vspace{2mm}
With the regression model in place, the selection between the CFD solver and the surrogate model is based on whether the predicted error and its associated uncertainty satisfy user-specified tolerances  $\tau_e$ and $\tau_u$. Specifically, we adopt a three-standard-deviation upper confidence bound, corresponding to a $99.7\%$ confidence level, and accept the surrogate model when
\begin{equation}
	3u_* \leq \tau_u,
\end{equation}	
and
\begin{equation}
	\mathbb{E}[{e}_*~\big|~\mathcal{A},\bm{e}] \leq \tau_e.
\end{equation}

When both criteria are satisfied, the surrogate model replaces the CFD solver and provides the fluid-induced load to the CSD solver for evaluating the structural design $\bm{\alpha}_*$. Otherwise, the framework reverts to the coupled CFD--CSD analysis \eqref{eq:discrete_coupled_problem_2}, and the resulting fluid solution is used to augment the dataset $\mathcal{D}$.

\vspace{2mm}
\noindent{\textit{Remark:}} For the GPR model to be well-defined, the equations \eqref{eq:beta_bar} and \eqref{eq:gp_variance} require the matrix $\bm{Y}\bm{K}^{-1}\bm{Y}^T \in \mathbb{R}^{N_\beta \times N_\beta}$ to be invertible. Since $\bm{K}$ is the covariance matrix with added measurement noise ($\sigma_m^2 > 0$), it is symmetric and positive definite, implying that $\bm{K}^{-1}$ exists and is also symmetric and positive definite. Using the Cholesky factorization of $\bm{K}^{-1}$ one can show that $\operatorname*{rank}(\bm{Y}\bm{K}^{-1}\bm{Y}^T) = \operatorname*{rank}(\bm{Y}) \leq \min(N_\beta, N_\text{true})$. Full rank, and hence invertibility, therefore requires $N_{\text{true}} \geq N_\beta$. For a polynomial basis of order $p$ in a $d$-dimensional design space, $N_\beta = \binom{d+p}{p}$, meaning at least this many high-fidelity CFD--CSD evaluations must be available before the decision model can be constructed.

\section{Implementation}\label{sec:implementation}

\subsection{Baseline optimization framework}\label{sec:baseline_optimization}

We begin by describing an implementation of the optimization process summarized in Algorithm~\ref{alg:gradient_free_fsi}, in which all design points are evaluated using coupled CFD--CSD analysis. This implementation serves as a baseline for the adaptive multi-fidelity framework, and is used to generate reference results for performance assessment. The framework consists of four major components that exchange data in real-time: a numerical optimizer, a mesh generator, a CFD solver, and a CSD solver.

\begin{figure}[H]
	\centering
	\includegraphics[width=0.8\linewidth]{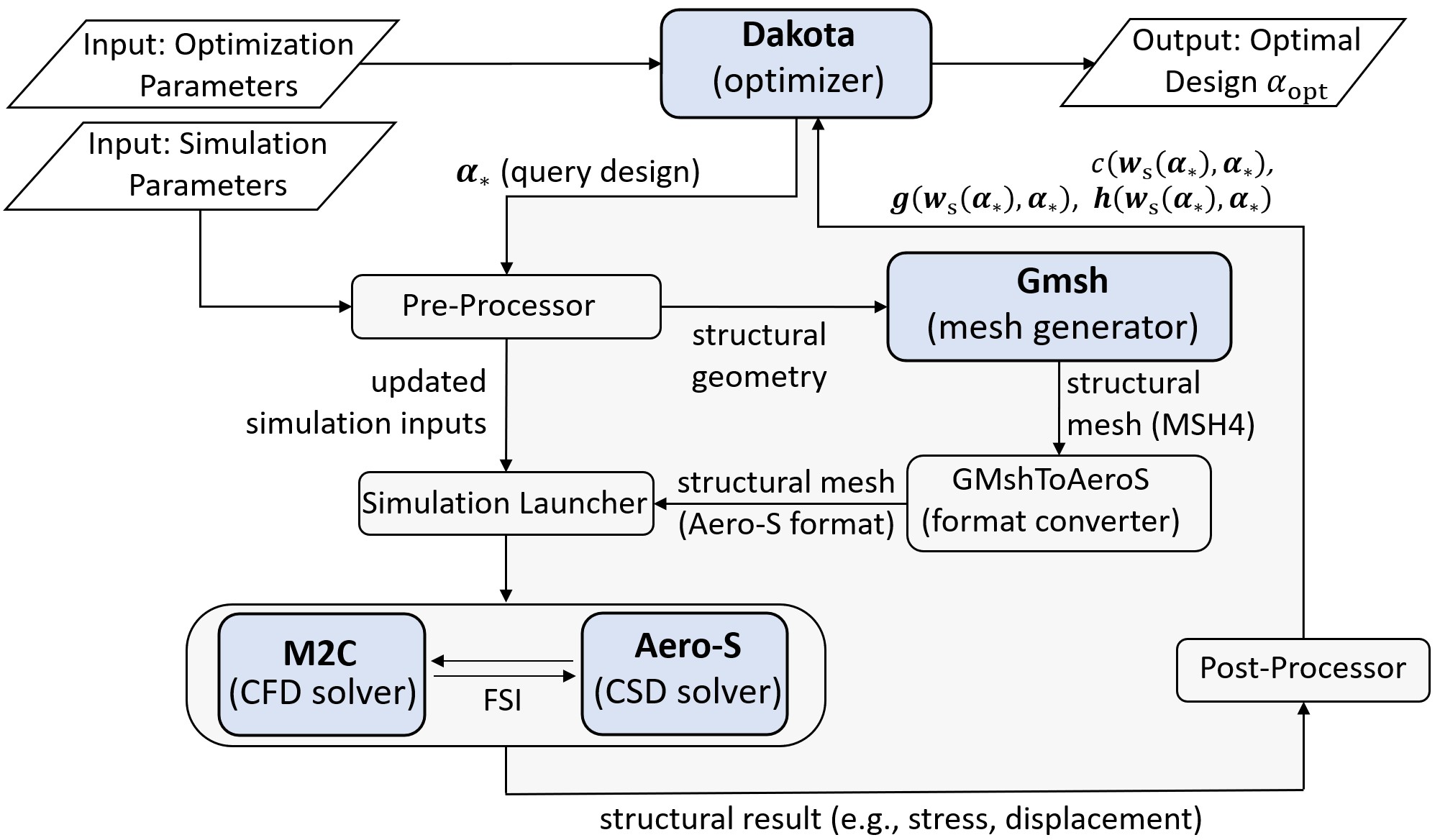}
	\caption{Implementation of the baseline optimization framework (SOFICS-Base).}
	\label{fig:sofics_v1_flowchart}
\end{figure}

Figure~\ref{fig:sofics_v1_flowchart} illustrates the workflow. The iterative optimization steps are shaded in gray, and the major components are highlighted in bold. The resulting software framework is named \emph{SOFICS} (Structural Optimization through Fluid–structure Interaction and Coupled Simulations). The optimization loop is driven by the open-source C++ toolkit Dakota (Design Analysis Kit for Optimization and Terascale Applications)~\cite{adams_dakota_2022}, which provides a variety of gradient-free algorithms  along with asynchronous and parallel evaluation capabilities. 

The fluid and structural sub-problems in \eqref{eq:discrete_coupled_problem_2} are solved using the CFD solver M2C and the CSD solver Aero-S, respectively. M2C is an Eulerian finite volume compressible flow solver, equipped with  an embedded boundary method for FSI~\cite{zhao2026m2c}. Aero-S is a Lagrangian finite element solver for structural and solid dynamics, supporting finite deformations and nonlinear material behaviors (e.g., plasticity, hyperelasticity, and viscoelasticity)~\cite{aeros}. The two solvers are both written in C++, and are coupled using a partitioned procedure and the FIVER (FInite Volume method based on Exact multi-material Riemann problems) method \cite{wang2011algorithms,main_enhanced_2017}. 

For each new structural design $\bm{\alpha}_*$, Dakota interfaces with the fluid and structural solvers through a Bash-based pre-processor. The pre-processor updates the simulation input files and calls the open-source meshing tool Gmsh~\cite{geuzaine_gmsh_2009} to create a new structural mesh. Because of the use of embedded boundary method, the fluid mesh can be the same for all design evaluations. Next, a simulation launcher executes the coupled analysis on a computer cluster. Upon completion, the results from Aero-S are post-processed to evaluate the objective function $c$ and the constraints $\bm{g}$ and $\bm{h}$.

The simulation launcher in SOFICS is designed to request a single allocation (i.e., fixed CPU cores and wall time) for the entire optimization process. Within this allocation, it runs coupled analysis concurrently and continuously using any available resources. This approach avoids delays and resource idleness associated with repeated queue submissions.

\subsection{Adaptive multi-fidelity framework}\label{sec:adaptive_framework}

The multi-fidelity optimization framework described in \sectionname~\ref{sec:adaptive_model} is implemented by extending the baseline optimization workflow, as summarized in Algorithm~\ref{alg:gradient_free_adaptive}. The design population management remains unchanged. The modifications primarily lie in the evaluation of individual candidate designs and the update of the decision model.

Each optimization iteration consists of four main steps. In Step~1, each candidate design is evaluated using either a coupled CFD--CSD analysis or a CSD analysis with fluid load provided by the surrogate model. The choice is determined by the decision model based on the predicted surrogate error and its uncertainty. If both quantities satisfy the prescribed tolerances, the surrogate model is used to approximate the fluid pressure, and only the structural sub-problem is solved. Otherwise, the coupled FSI problem is solved, and the design is added to the set $\mathcal{A}$ of designs with available high-fidelity data. In either case, the resulting structural state is used to evaluate the objective function and constraints. In Step~2, the best design in the current population is identified and the stopping criterion is checked. If the optimization continues, Step~3 updates the surrogate error dataset and GPR-based decision model using the available high-fidelity data. Finally, in Step~4, the gradient-free optimization algorithm generates the candidate designs for the next iteration. These four steps are repeated until the stopping criterion is satisfied or the maximum number of iterations is reached. Compared to the baseline algorithm (Algorithm~\ref{alg:gradient_free_fsi}), the multi-fidelity framework introduces three additional user-specified parameters, namely the number of neighbors $N_b$ used in the surrogate model of fluid-induced loads (Sec.~\ref{sec:pressure_approx}), and the tolerances $\tau_e$ and $\tau_u$ used in the adaptive decision model (Sec.~\ref{sec:adaptive_model}). The other parameters, including the interpolation scaling factor in \eqref{eq:gaussian_rbf} and the GPR parameters in \eqref{eq:hyperparameters}, are determined from the data accumulated during the optimization.

\begin{algorithm}[H]
	\caption{Adaptive Multi-Fidelity Structural Optimization under Fluid-Structure Interaction}
	\label{alg:gradient_free_adaptive}
	\begin{algorithmic}[1]
		\Require Initial design candidates ${\bm{\alpha}^{(0)}_1, \dots, \bm{\alpha}^{(0)}_M}$, maximum optimization iteration $N_\text{iter}$, optimization stopping criterion (e.g., change in objective function's value), penalty factory $\mu$, tolerances $\tau_e$ and $\tau_u$, and number of neighbors $N_b$
		
		\State Set $\mathcal{D} = \varnothing$, $\mathcal{A} = \varnothing$.
		
		\For{$k = 1, \dots, N_{\text{iter}}$}
				
		\vspace{2mm}
		\State Set $N_\text{true} = \lvert\mathcal{A}\rvert $.
		\vspace{2mm}
		
		\State \textbf{Step 1: Multi-fidelity analysis}
		
		\For{$i = 1, \dots, M$}
		
		\State $\bm{\alpha}_* \gets \bm{\alpha}_i^{(k)}$
		
		\State $\texttt{accept\_surrogate} \gets \texttt{false}$
		
		\If{$N_\text{true}  \geq N_\beta$}
		
		\State Compute $\bm{y}_*$, $\bm{k}_*$, and $k_{**}$ as defined in Sec.~\ref{sec:adaptive_model}.
		
		\State Evaluate $\mathbb{E}\big[e_*~|~\mathcal{A},\bm{e}\big]$ by \eqref{eq:gp_expectation} and $u_*$ by \eqref{eq:gp_variance} and \eqref{eq:uncertainty}.
		
		\Comment{Use the latest $\bm{K}^{-1}$, $\bm{Y}$, $\bar{\bm{\beta}}$, and $\bm{e}$ computed in {\bf Step 3}} 
		
		\If{$3u_* \leq \tau_u$ \textbf{and} $\mathbb{E}\big[e_*~|~\mathcal{A},\bm{e}\big] \leq \tau_e$}
		\State $\texttt{accept\_surrogate} \gets \texttt{true}$
		\EndIf
		\EndIf
		
		\vspace{1mm}
		
		\If{$\texttt{accept\_surrogate}$}
		\State Identify nearest neighbors $\{\bm{\alpha}_{1}, \ldots, \bm{\alpha}_{N_b}\} \subset \mathcal{A}$ of $\bm{\alpha}_*$ such that,\vspace{2mm}
		\State \quad\quad $\norm{\bm{\alpha}_* - \bm{\alpha}_{1}} \leq \cdots \leq \norm{\bm{\alpha}_* - \bm{\alpha}_{N_b}} \leq \norm{\bm{\alpha}_* - \bm{\alpha}'}, \quad \forall \bm{\alpha}' \in \mathcal{A} \setminus \{\bm{\alpha}_{1}, \ldots, \bm{\alpha}_{N_b}\}$.\vspace{2mm}
		
		\State Solve structural sub-problem $\mathcal{R}_\text{S}(\tilde{p}, \bm{w}_\text{S}; \bm{\alpha}_*) = \bm{0}$ for $\bm{w}_\text{S}$ with  boundary condition 
		\vspace{2mm}
		\State \quad\quad $\phi_2 = \big(\boldsymbol{\sigma}_\text{S} - \tilde{p}\big(\bm{X},~t;~\bm{\alpha}_*\big) \mathbb{I}\big)\cdot \boldsymbol{n} = \bm{0}$,\quad\quad(cf.~\eqref{eq:dynamic_interf_cond})\vspace{2mm}
		\State where $\tilde{p}$ is computed using the surrogate model \eqref{eq:pressure_formula}.
		\Else
		\State Solve coupled FSI problem $\mathcal{R}(\bm{w}_\text{F}, \bm{w}_\text{S}; \bm{\alpha}_*) = \bm{0}$ for $\bm{w}_\text{F}$ and $\bm{w}_\text{S}$.
		\State $\mathcal{A} \gets \mathcal{A} \cup \left\{\bm{\alpha}_*\right\}$
		\EndIf
		\State Evaluate objective function $c(\bm{w}_\text{S}, \bm{\alpha}_*)$ and constraints $\bm{g}(\bm{w}_\text{S}, \bm{\alpha}_*)$, $\bm{h}(\bm{w}_\text{S}, \bm{\alpha}_*)$.
		\EndFor
		
		\vspace{2mm}
		\State \textbf{Step 2: Find best design and check stopping criterion}
		\State Identify the best design in the current iteration:\vspace{2mm}
		\State \quad\quad $\bm{\alpha}_{\text{opt}} = \underset{\left\{\bm{\alpha}_i^{(k)},~i=1,\dots,M\right\}}{\arg\min}~\left[ c(\bm{w}_\text{S}, \bm{\alpha}_i^{(k)}) + \mu \left( \norm{\max(\bm{0}, \bm{g}(\bm{w}_\text{S}, \bm{\alpha}_i^{(k)}))}^2 + \norm{\bm{h}(\bm{w}_\text{S}, \bm{\alpha}_i^{(k)})}^2 \right) \right]$\vspace{2mm}
		
		\If{$\bm{\alpha}_{\text{opt}} \notin \mathcal{A}$}
		\State Solve coupled FSI problem $\mathcal{R}(\bm{w}_\text{F}, \bm{w}_\text{S}; \bm{\alpha}_{\text{opt}}) = \bm{0}$ for $\bm{w}_\text{F}$ and $\bm{w}_\text{S}$.
		\State $\mathcal{A} \gets \mathcal{A} \cup \{\bm{\alpha}_{\text{opt}}\}$
		\State Re-evaluate objective function $c(\bm{w}_\text{S}, \bm{\alpha}_{\text{opt}})$ and constraints $\bm{g}(\bm{w}_\text{S}, \bm{\alpha}_{\text{opt}})$, $\bm{h}(\bm{w}_\text{S}, \bm{\alpha}_{\text{opt}})$.
		\EndIf

		\algstore{adaptiveAlg}
	\end{algorithmic}
\end{algorithm}

\clearpage

\begin{algorithm}[!htbp]
	\begin{algorithmic}[1]
		\algrestore{adaptiveAlg}
		\State Check stopping criterion. If satisfied,
		\State \quad\quad \textbf{break}		

\vspace{2mm}

		\State \textbf{Step 3: Update dataset and GPR model}
		\State Update $N_{\text{true}} = \lvert\mathcal{A}\rvert$.
		\State Reset $\mathcal{D} = \varnothing$.
		\If{$N_\text{true} \geq N_\beta$}
		\For{$i = 1, \dots, N_{\text{true}}$}
		\State Compute $\tilde{p}(\bm{X},t;\bm{\alpha}_i)$ by \eqref{eq:pressure_formula}, using $N_b$ nearest neighbors from $\mathcal{A} \setminus \{\bm{\alpha}_i\}$.
		\State Compute $e_i = e(\bm{\alpha}_i)$ by \eqref{eq:rms_error}.
		\State $\mathcal{D} \gets \mathcal{D} \cup \left\{(\bm{\alpha}_i,e_i)\right\}$
		\EndFor
		\State Assemble vector $\bm{e} = \left[e_1,e_2,\dots,e_{N_\text{true}}\right]^T$ and matrix $\bm{Y} = \left[\bm{y}(\bm{\alpha}_1),\bm{y}(\bm{\alpha}_2),\dots,\bm{y}(\bm{\alpha}_{N_\text{true}})\right]$.
		\State Compute $\{\sigma_m^2, \{\theta_j\}_{j=1}^d, \sigma^2\}$ by maximizing log marginal likelihood \eqref{eq:hyperparameters}.
		\State Assemble covariance matrix $\bm{K} \in \mathbb{R}^{N_{\text{true}} \times N_{\text{true}}}$ by \eqref{eq:K_entries}; compute $\bm{K}^{-1}$ by Cholesky factorization.
		\State Compute $\bar{\bm{\beta}}$ by \eqref{eq:beta_bar}.
		\EndIf
		
		\vspace{2mm}
		\State \textbf{Step 4: Update design candidates}
		\State Apply a gradient-free update rule to obtain new design candidates $\{\bm{\alpha}^{(k+1)}_1, \dots, \bm{\alpha}^{(k+1)}_M\}$.
		\EndFor
		
		\vspace{2mm}
		\Ensure Final design: $\bm{\alpha}_\text{opt}$
	\end{algorithmic}
\end{algorithm}


A few details in Algorithm~\ref{alg:gradient_free_adaptive} deserve further clarification. During the first iteration ($k=1$), and until the number of coupled CFD--CSD analyses is sufficiently large to construct the GPR model (i.e., $N_{\text{true}} \ge N_{\bm{\beta}}$), all candidate designs are evaluated using coupled simulations. Once the GPR model has been constructed, Lines~9 and 10 use the model obtained from Step~3 of the previous iteration to predict the surrogate error and its uncertainty for each new candidate design. As additional coupled analyses are performed in Step~1, the corresponding designs are added to $\mathcal{A}$ and become available for subsequent surrogate and decision-model updates.

In Step~2, if the best design in the current iteration was evaluated using the surrogate model, an additional coupled CFD--CSD analysis is performed at that design point. The objective function and constraints are re-evaluated using its result. In this way, the design used to check the stopping criterion and ultimately returned by the optimizer is always evaluated using a coupled  FSI analysis. In Step~3, the error dataset $\mathcal{D}$ is recomputed at every iteration. For each design in $\mathcal{A}$, the surrogate prediction is reconstructed using its $N_b$ nearest neighbors, and the corresponding surrogate error is evaluated. The resulting error data are then used to update the  GPR model and its hyperparameters. This recomputation is necessary because the addition of high-fidelity data changes the surrogate model and, consequently, its errors at previously evaluated design points.

Although not shown explicitly in the algorithm, the fluid pressure and structural displacement fields obtained from each coupled CFD--CSD analysis are stored throughout the optimization process, as they are used by the surrogate model. While Algorithm~\ref{alg:gradient_free_adaptive} assumes that the fluid-induced loads are pressure loads, it can be readily extended to handle the more general case, where the interface traction includes both normal and shear components.
\vspace{2mm}

\begin{figure}[H]
\centering
\includegraphics[width=0.8\linewidth]{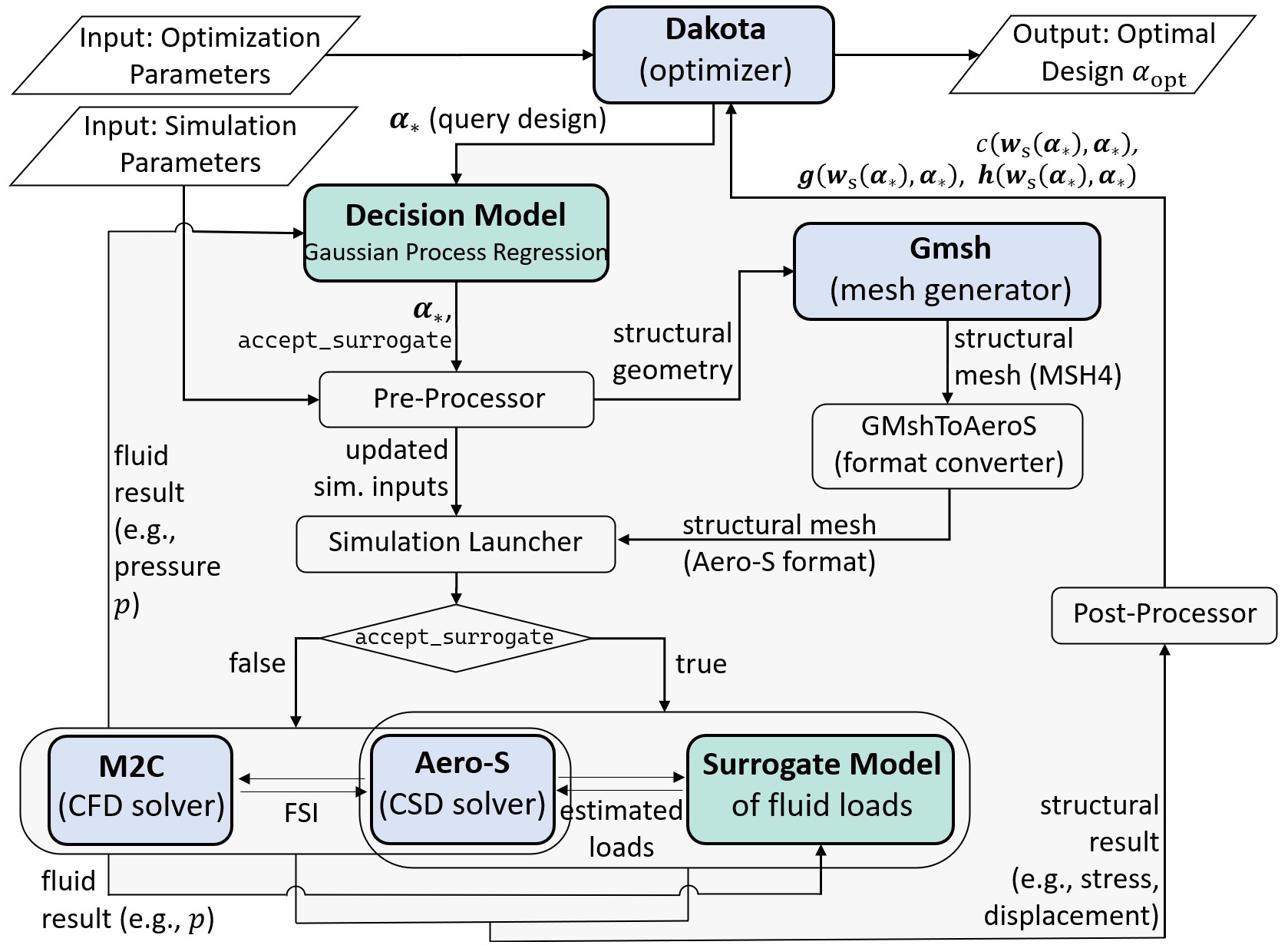}
\caption{Implementation of the adaptive multi-fidelity optimization framework (SOFICS-AMF).}
\label{fig:sofics_v2_flowchart}
\end{figure}

Fig.~\ref{fig:sofics_v2_flowchart} illustrates our implementation of Algorithm~\ref{alg:gradient_free_adaptive}. An important consideration in this implementation is software modularity. Compared with the baseline architecture shown in Fig.~\ref{fig:sofics_v1_flowchart}, the two new components are the surrogate model for fluid-induced loads and the GPR-based decision model. In contrast, the Dakota optimizer, the M2C solver, the Aero-S solver, and the Gmsh mesh generator require no modifications.

The surrogate model is implemented to follow the same communication and data exchange routines as those implemented in M2C. As a result, Aero-S does not need to know whether it is communicating with M2C or the surrogate model. The pre-processing and simulation launcher are modified slightly to accept the additional input flag \texttt{accept\_surrogate}. 

This new architecture is referred to as SOFICS-AMF, to distinguish it from the baseline SOFICS-Base illustrated in Fig.~\ref{fig:sofics_v1_flowchart}. However, SOFICS-AMF can be viewed as a backward-compatible extension of SOFICS-Base. Setting the decision model tolerances $\tau_u=\tau_e=0$ reduces the framework to the baseline workflow, in which the surrogate model is never accepted and all design evaluations are performed through coupled analyses.

\subsection{Computational cost}\label{sec:adaptive_computational_cost}

While the adaptive decision model introduces the GPR into the workflow, the associated computational overhead is trivial compared to the cost of coupled CFD--CSD analysis.

In the worst-case scenario, the GPR model decides to evaluate all candidate designs using coupled CFD--CSD analyses. The additional overhead incurred comes from three operations: (1) computing the fluid pressure (or traction vector) approximation errors, (2) updating the error dataset $\mathcal{D}$, and (3) calibrating the GPR model using the updated dataset. Among these, the error computation step is inexpensive, as it only involves evaluating quantities defined on the lower-dimensional fluid-structure interface rather than the full fluid domain $\Omega_\text{F}$. Additionally, the nearest-neighbor search required for pressure approximation has a computational complexity of $\mathcal{O}(d N_\text{true})$, which remains modest because both the design space dimension $d$ and the number of FSI samples $N_\text{true}$ are small relative to the size of the full fluid discretization.

The dominant additional cost comes from calibrating the GPR model, which may scale cubically with the size of the regression dataset, i.e., $\mathcal{O}(N_\text{true}^3)$. However, this is still trivial compared to the cost of a coupled CFD--CSD analysis. In contrast, the cost of a coupled CFD--CSD analysis is typically on the order of $\mathcal{O}(N^2 \log{N})$ for the fluid sub-problem, where $N$ represents the total number of spatial and temporal grid points, and on the order of $\mathcal{O}(n^2 \log n)$ for the structural sub-problem, with $n \ll N$. As a result, unless the size of the GPR training dataset approaches the scale of the fluid discretization---which is rarely the case in practice---the overhead introduced by the adaptive framework remains negligible relative to the cost of solving a FSI problem.

\section{Numerical Experiments}\label{sec:results}

\subsection{Piston in one-dimensional flow}\label{sec:piston}

\subsubsection{Problem definition}
\label{sec:piston_def}

\begin{figure}[H]
	\centering
	\includegraphics[width=1\linewidth]{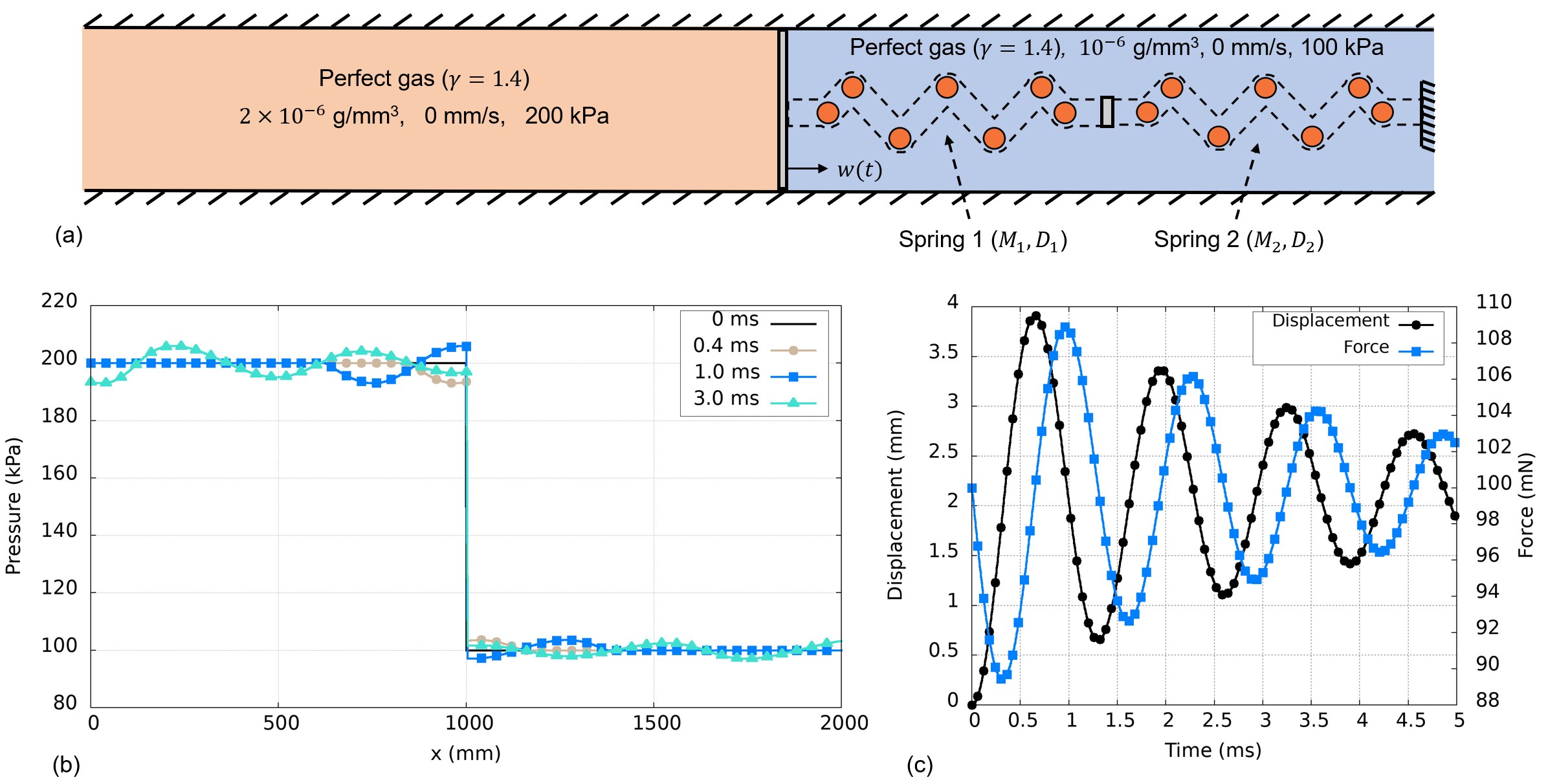}
	\caption{Piston in a one-dimensional flow. (a) Problem setup. (b) Fluid pressure field for $\alpha_{(1)} = 1.25$ and $\alpha_{(2)} = 0.75$. (c) Time histories of piston displacement and fluid load, $p_{\mathrm{F}}(t)A$, for the same design.}
	\label{fig:piston_model}
\end{figure}

We consider a one-dimensional FSI problem involving a compressible, inviscid fluid flow interacting with a piston supported by two linear springs. The problem setup is illustrated in Fig.~\ref{fig:piston_model}(a). The fluid sub-problem, $\mathcal{R}_{\text{F}}$, is governed by Eqs.~\eqref{eq:fluid_coupled_equation} and \eqref{eq:euler_terms}, reduced to one spatial dimension. The structural sub-problem, $\mathcal{R}_{\text{S}}$, is described by
\begin{equation}
	\bar{M} \ddot{w}(t) + \bar{D} w(t) = p_{\text{F}}(t)A,\quad
	w(0) = 0,\quad
	\dot{w}(0) = 0,
	\label{eq:piston_1d}
\end{equation}
where $w(t)$ denotes the piston displacement. $\bar{M}$ and $\bar{D}$ are the structure's mass and stiffness, given by
\begin{equation}
	\bar{M} = M_1 + M_2,\qquad
	\bar{D} = \left(\dfrac{1}{D_1} + \dfrac{1}{D_2}\right)^{-1},
\end{equation}
with $M_i$ and $D_i$ representing the mass and stiffness of spring $i$ ($i=1,2$). Here, $p_{\text{F}}(t)$ denotes the fluid pressure acting on the piston surface, and $A = 1~\text{mm}^2$ is the piston area.

We introduce a two-dimensional design space parameterized by the non-dimensional design variables $\bm{\alpha} = [\alpha_{(1)},~\alpha_{(2)}]^T$, which controls both the mass and stiffness of each spring. Specifically, for $i=1,2$,
\begin{equation}
	D_i = c_D \alpha_{(i)},\qquad
	M_i = c_M \alpha_{(i)},
	\label{eq:spring_coeffs}
\end{equation}
where $c_D = 10^5~\text{g}/\text{s}^2$ and $c_M = 10^{-3}~\text{g}$. 

The optimization problem seeks to minimize the maximum piston velocity, $\|\dot{w}(t)\|_{\infty}$, subject to inequality constraints on the design variables and piston displacement:
\begin{equation}
	\begin{aligned}
		\underset{\boldsymbol{\alpha} \in \Omega_D}{\operatorname*{minimize}} \quad
		& \|\dot{w}(t;\boldsymbol{\alpha})\|_{\infty} \\
		\text{subject to} \quad
		& \alpha_{(1)} + \alpha_{(2)} \leq 2.0, \\
		& \alpha_{(1)} - \alpha_{(2)} \geq 0.5, \\
		& w(t;\boldsymbol{\alpha}) \leq 10.0~\text{mm}, \quad \forall t > 0,
	\end{aligned}
	\label{eq:piston_opt_problem}
\end{equation}
where $\Omega_D = [0.2,~2.0]^2$.

By construction, the optimal solution occurs when the first two constraints hold as equalities, yielding $\boldsymbol{\alpha}_{\text{opt}} = [1.25,~0.75]^T$. This can be understood by noticing that the maximum displacement and velocity both decrease with increasing mass $\bar{M}$ and stiffness $\bar{D}$. When $\bar{M} = c_M(\alpha_{(1)} + \alpha_{(2)})$ is fixed, $\bar{D}$ is maximized when $\alpha_{(1)}$ and $\alpha_{(2)}$ are as close as possible.

\subsubsection{Optimization results}
\label{sec:piston_opt_res}

Figure~\ref{fig:piston_model}(b) and (c) show the results of the coupled FSI analysis at the optimal design. Initially, a higher fluid pressure on the left side drives the piston to the right. As the piston compresses, the restoring elastic force increases and eventually reverses the motion, pushing the piston back to the left, ultimately leading to an oscillatory motion.

The optimization problem is solved using the two methods described in Sec.~\ref{sec:implementation}:
\begin{itemize}
	\item \textbf{BASE}: The baseline  framework from Sec.~\ref{sec:baseline_optimization} (i.e.,~SOFICS-Base) is employed. Every candidate design is evaluated using a coupled CFD--CSD analysis. The piston surface is represented as an embedded surface in the fluid domain, and the same fluid mesh is used for all designs. 	
	\item \textbf{AMF}: The adaptive multi-fidelity framework from Sec.~\ref{sec:adaptive_framework} (i.e.,~SOFICS-AMF) is used. For each candidate design, the GPR model predicts the surface pressure approximation error and its associated uncertainty. Using the tolerances $\tau_e = 1.3\%$ and $\tau_u = 1\%$, the decision model selects between a coupled CFD--CSD analysis and a structure-only analysis in which transient, distributed fluid pressure loads are computed using surrogate model \eqref{eq:pressure_formula}. The pressure is interpolated from $N_b = 5$ nearest neighbors in the design space.
\end{itemize}

The same genetic algorithm settings are used in both cases. The initial population consists of $20$ designs generated randomly within $\Omega_D$. The selection operator is chosen to favor feasible designs. Shuffle-random crossover is used with a crossover rate of $0.1$, while offset-normal mutation is used with a mutation rate of $1.0$ and a mutation scale of $0.02$. In both cases, the optimization is run for $20$ iterations.

\begin{figure}[H]
	\centering
	\includegraphics[width=1\linewidth]{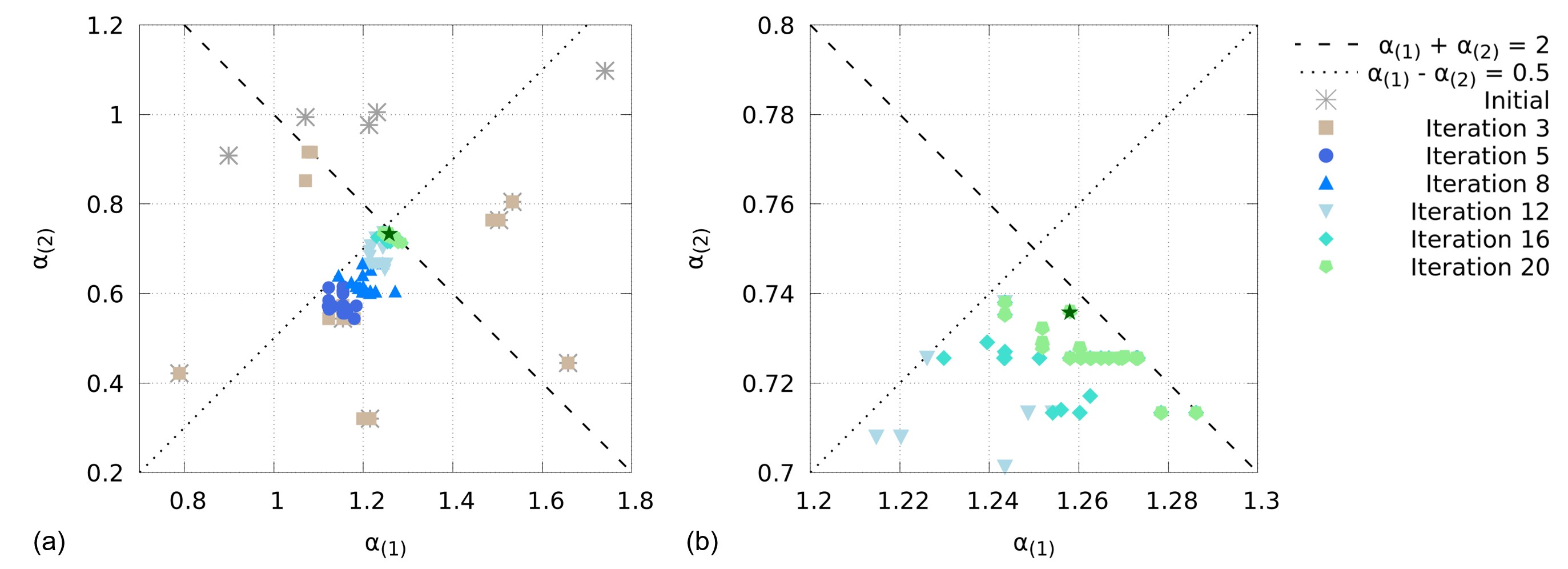}
	\caption{Evolution of the design populations in \textbf{BASE}. (a) Broader view of the design space; (b) detailed view in the vicinity of the optimum. The final design is marked with a pentagram. For reference, the two linear constraints are also shown, and the true optimum lies at their intersection.}
	\label{fig:piston_base}
\end{figure}

Figure~\ref{fig:piston_base} presents the evolution of the design populations in \textbf{BASE}. The two linear inequality constraints active at the optimum are also plotted. Beginning at Iteration 5, the population lies entirely within the feasible region and progressively converges toward the true solution. At the final iteration (Iteration $20$), the optimizer provides  $\bm{\alpha}_{\text{opt}}^{\text{(BASE)}} = [1.258,~0.736]^T$, which differs from the true optimum by $1.1\%$.

\begin{figure}[H]
	\centering
	\includegraphics[width=1\linewidth]{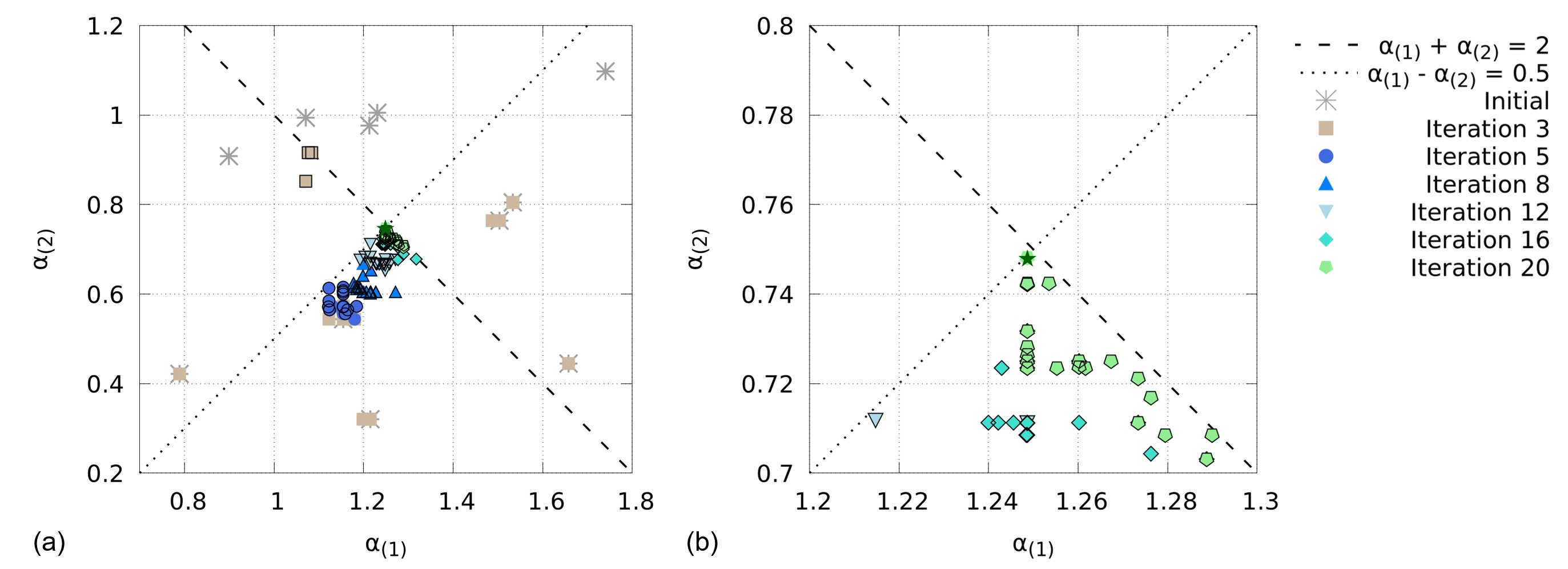}
	\caption{Evolution of the design populations in \textbf{AMF}. (a) Broader view of the design space; (b) detailed view in the vicinity of the optimum. Designs evaluated using the surrogate model are indicated by black outlines, and the final design is marked with a pentagram.}
	\label{fig:piston_amf}
\end{figure}

Figure~\ref{fig:piston_amf}  shows the corresponding results for \textbf{AMF}. Similar convergence toward the true optimum is observed throughout the optimization process. At Iteration $20$, the final design is $\bm{\alpha}_{\text{opt}}^{\text{(AMF)}} = [1.249,~0.748]^T$, corresponding to an error of only $0.16\%$. While the convergence trend is expected, the smaller final error relative to \textbf{BASE} is likely coincidental.

Figure~\ref{fig:piston_amf} identifies the designs evaluated using the surrogate fluid model with black outlines. Beginning in Iteration 3, three designs are evaluated using the surrogate model. This number increases to $13$ in Iteration 5, and from Iteration 6 onward, the vast majority of designs are evaluated in this way. Overall, $347$ out of $407$ design evaluations ($85\%$) are performed using the surrogate model. By comparison, \textbf{BASE} also requires $407$ design evaluations, all based on coupled analyses. Since the computational cost of the piston solver is negligible relative to the CFD solver, the overall computational cost reduction is close to $85\%$.

\noindent{\it Remark:} This example also illustrates the need of the hybrid Lagrangian-Eulerian mapping introduced in Sec.~\ref{sec:pressure_approx}. As shown in Fig.~\ref{fig:piston_model}(b), the fluid pressure is discontinuous across the piston, with the pressures on  the left and right sides oscillating around $200$ and $100$ kPa, respectively. Therefore, when approximating the pressure load on either side of the piston, the interpolation data must be taken from the corresponding fluid subdomain. Because the piston moves over time, interpolation based solely on spatial proximity in the deformed configuration could inadvertently use pressure data from the opposite side of the piston.  The hybrid Lagrangian-Eulerian mapping preserves this correspondence, thereby avoiding the use of fluid pressure from the wrong side of the structure.

\subsubsection{Robustness assessment}


First, we assess the sensitivity of the proposed adaptive multi-fidelity framework to the number of neighbors $N_b$ in the fluid surrogate model. The \textbf{AMF} optimization process is repeated five times for $N_b=1,\ldots,5$, with the main results summarized in Table~\ref{tab:amf_ablation}. The five studies return merit function values between $0.3160$ and $0.3181$, corresponding to a spread of $0.66\%$, with no systematic dependence on $N_b$. These values are directly comparable across studies because the best design of each iteration is re-evaluated using coupled CFD--CSD analysis before the optimization proceeds (Sec.~\ref{sec:adaptive_framework}). Therefore, the reported merit function values do not contain surrogate approximation errors. These results suggest that, for this problem and the range considered, the optimization outcome is relatively insensitive to the choice of $N_b$.

\begin{table}[H]
	\centering
	\begin{tabular}{lcccc}
		\toprule
		$N_b$ & $c(\bm{w}_\text{S}, \bm{\alpha}^\text{opt})$ & Coupled \,/\, Total evaluations & Coupled evaluations (\%) & False negatives\\
		\midrule
		$1$ & $0.3181$ & $31$ \,/\, $414$ & $7.5\%$   &  $5$\\
		$2$ & $0.3160$ & $43$ \,/\, $416$ & $10.3\%$  &  $5$\\
		$3$ & $0.3173$ & $39$ \,/\, $415$ & $9.4\%$   &  $6$\\
		$4$ & $0.3168$ & $48$ \,/\, $413$ & $11.6\%$  &  $1$\\
		$5$ & $0.3172$ & $61$ \,/\, $414$ & $14.7\%$  &  $2$\\
		\bottomrule
	\end{tabular}
	\caption{Effect of the number of neighbors $N_b$ used in the surrogate model on the optimization outcome, computational cost, and false-negative decisions.}
	\label{tab:amf_ablation}
\end{table}

Table~\ref{tab:amf_ablation} also shows the total number of design evaluations and the number performed using coupled CFD--CSD analysis. The total number of evaluations is nearly identical among the five cases, whereas the number of coupled analyses increases from $7.5\%$ for $N_b=1$ to $14.7\%$ for $N_b=5$. This trend can be explained by the use of the interpolation database. Because the database is sparse during the early optimization iterations, a larger neighbor set tends to include designs farther from the query point $\bm{\alpha}_*$. This generally increases the measured errors $e$ recorded in $\mathcal{D}$ and, consequently, leads the decision model to accept the surrogate less frequently.

\vspace{2mm}

We assess the decision model by identifying cases in which the surrogate fluid model was accepted even though its actual approximation error exceeded the prescribed tolerance $\tau_e$. Such cases are referred to here as ``false negatives.'' For design points where the surrogate model is accepted, the actual approximation error is not available during optimization because the corresponding coupled CFD--CSD analysis is not performed. We  recover this information \textit{a posteriori} by re-evaluating every such design using coupled analysis. Then, the actual error $e(\bm{\alpha}_*)$ is computed using \eqref{eq:rms_error}.

Across the five studies, a total of $1{,}848$ surrogate-assisted design evaluations are re-evaluated in this manner. As reported in Table~\ref{tab:amf_ablation}, $19$ ($1.0\%$) are identified as false negatives. The number ranges from one to six among the five values of $N_b$, with no apparent trend. These cases are not associated with large predicted uncertainty: for all $19$ false negatives, $3u_*$ ranges from $4.6\times10^{-4}$ to $5.4\times10^{-3}$. Instead, they can be understood from the assumptions underlying the GPR model. The posterior variance \eqref{eq:gp_variance} reflects, in part, how well the design space is sampled near a query point: a small value generally corresponds to interpolation among nearby data, whereas a larger value often implies extrapolation. It does not, however, directly measure the accuracy of the posterior mean \eqref{eq:gp_expectation}. Furthermore, the decision model employs a covariance function that relies on a single matrix $\bm{\Theta}$ of inverse length scales for the entire design space, resulting in a smooth posterior mean. A false negative can therefore occur when the data are locally dense, leading to low predicted uncertainty, while the actual error field varies more rapidly than the stationary covariance model can represent.

For example, for $N_b=1$ and at iteration $3$, the design $\bm{\alpha}_*=[1.215,\ 0.251]^T$ has its two nearest neighbors  in $\mathcal{A}$ at a distance of approximately $0.070$, both with $e\approx1.24\times10^{-3}$. Based on the available data, the GPR model at this iteration estimates $\mathbb{E}[e_*\,|\,\mathcal{A},\bm{e}]=9.1\times10^{-4}$, whereas the true error is $2.55\times10^{-2}$. Thus, the actual error increases by approximately a factor of twenty over a distance for which the GPR model (at that iteration) varies by less than a factor of two. Because the correlation length obtained from \eqref{eq:hyperparameters} is estimated from the entire database, it appears to be too large to resolve this local variation. This example illustrates a limitation of using a stationary covariance function when the characteristic length scale of the error field varies across the design space.

\vspace{2mm}

Furthermore, we compare the GPR-based decision model with a simpler fidelity selection criterion based on radial basis interpolation of surrogate error. Instead of using GPR, we estimate the surrogate approximation error \eqref{eq:rms_error} at each query design $\bm{\alpha}_*$ by
\begin{equation}
e_\text{rbf}(\bm{\alpha}_*) = \sum_{k=1}^{N_r} w_k e(\bm{\alpha}_k),
\label{eq:rbf_error}
\end{equation}
where
\begin{equation}
w_k =
	\sum_{m=1}^{N_r}
	\phi\!\left(\lVert \bm{\alpha}_* - \bm{\alpha}_m \rVert\right)
		\left(\bm{\Phi}^{-1}\right)_{mk}
\end{equation}
denotes the interpolation weight for neighbor $\bm{\alpha}_k$.  $\bm{\Phi}$ is the interpolation matrix obtained with Gaussian radial basis function \eqref{eq:gaussian_rbf}, with scaling factor $l$ set to the average distance between the neighboring designs in $\mathcal{A}$. Unlike the GPR model, this method does not provide a measure of uncertainty. The fluid-load surrogate model is therefore accepted whenever $e_\text{rbf}(\bm{\alpha}_*) \le \tau_e$. Here, $\tau_e$ is set to $1.3\%$, consistent with the value used in the GPR-based decision model.

The five studies reported in Table~\ref{tab:amf_ablation} are repeated using the simpler fidelity selection model, with $N_r=5$ in all cases.  Table~\ref{tab:amf_ablation2} summarizes the main results. The merit function value varies between $0.3161$ and $0.3173$, comparable to those obtained using the GPR model. Following the same \textit{a posteriori} procedure described above, $11$ of the $1{,}446$ surrogate-assisted evaluations ($0.8\%$) are identified as false negatives, compared with $1.0\%$ for the GPR model. 

\begin{table}[H]
	\centering
	\begin{tabular}{lcccc}
		\toprule
		$N_b$ & $c(\bm{w}_\text{S}, \bm{\alpha}^\text{opt})$ & Coupled \,/\, Total evaluations & Coupled evaluations (\%) & False negatives\\
		\midrule
		$1$ & $0.3161$ & $29$ \,/\, $416$  & $7.0\%$   &  $5$\\
		$2$ & $0.3165$ & $194$ \,/\, $408$ & $47.5\%$  &  $0$\\
		$3$ & $0.3161$ & $103$ \,/\, $410$ & $25.1\%$  &  $1$\\
		$4$ & $0.3173$ & $135$ \,/\, $410$ & $32.9\%$  &  $1$\\
		$5$ & $0.3166$ & $145$ \,/\, $408$ & $35.5\%$  &  $4$\\
		\bottomrule
	\end{tabular}
	\caption{Optimization results obtained using a simpler interpolation-based decision model.}
	\label{tab:amf_ablation2}
\end{table}

The computational cost, however, differs substantially. For $N_b = 2,~3,~4,$ and $5$, the proportion of designs evaluated by coupled CFD--CSD analysis increases from $9.4$--$14.7\%$ to $25.1$--$47.5\%$. This increase can be explained by the conditioning of \eqref{eq:rbf_error} as the optimization proceeds. 
As the optimization converges, the design points used in the interpolation become increasingly clustered, producing nearly identical rows in matrix $\bm{\Phi}$.  In comparison, the GPR model is not subject to this issue as the measurement variance $\sigma_m^2$ added to the diagonal of the covariance matrix $\bm{K}$ in \eqref{eq:K_entries} keeps it well-conditioned. 

\subsubsection{Scalability analysis}


We examine how the framework behaves as the dimension of the design space increases. The fluid sub-problem, fluid mesh, and the structural model equation~\eqref{eq:piston_1d} remain unchanged. The spring assembly is extended to include a variable number of springs, denoted by $d$. The design vector $\bm{\alpha}=[\alpha_{(1)},\dots,\alpha_{(d)}]^T$ contains one non-dimensional variable per spring, with the mass and stiffness of each spring given by \eqref{eq:spring_coeffs} and $\Omega_D=[0.2,\,2.0]^d$. The springs remain connected in series, such that
\begin{equation}
	\bar{M}=c_M\sum_{i=1}^{d}\alpha_{(i)}, \qquad
	\bar{D}=c_D\left(\sum_{i=1}^{d}\alpha_{(i)}^{-1}\right)^{-1}.
\end{equation}

The linear constraints in the optimization are now applied to consecutive pairs of design variables. That is, for $j=1,\dots,d/2$,
\begin{equation}
	\alpha_{(2j-1)}+\alpha_{(2j)}\leq 2.0, \qquad
	\alpha_{(2j-1)}-\alpha_{(2j)}\geq 0.5 .
\end{equation}
Thus, each pair of springs is subject to the same constraints analyzed in Sec.~\ref{sec:piston}, and the feasible region decouples by pair. This construction allows the design space to be enlarged while preserving the basic characteristics of the optimization problem and its solution. The optimum for each pair lies at the intersection of its two constraints. Therefore, the exact optimum is the two-variable solution from Sec.~\ref{sec:piston_def} repeated $d/2$ times, i.e., $\bm{\alpha}_{\text{opt}}=[1.25,~0.75]^T$ for $d=2$, $[1.25,~0.75,~1.25,~0.75]^T$ for $d=4$, and so forth.

Four studies are conducted with $d=2$, $4$, $6$, and $8$. The genetic algorithm settings, decision model tolerances ($\tau_e=1.3\%$, $\tau_u=1\%$), and number of neighbors ($N_b=5$) are fixed across all studies. The dimension of the design space is the only parameter varied.

Because the number of iterations required for convergence grows with $d$, terminating all four studies after a prescribed number of iterations, as done in Sec.~\ref{sec:piston_opt_res}, would not reflect the computational effort required to solve the optimization problem. Therefore, we use a stopping criterion based on the exact solution. For each study, a coupled CFD--CSD analysis is first performed at $\bm{\alpha}_{\rm opt}$ to obtain the corresponding merit function value $c_{\rm opt}$. During the optimization process, the best design in each iteration is compared with this reference, and the optimization is terminated once its merit function value is within $2\%$ of $c_{\rm opt}$. This criterion allows \textbf{BASE} and \textbf{AMF} to be compared at the same level of solution quality, with the computational effort required to reach that level serving as the measure of cost. 

\begin{figure}[H]
	\centering
	\includegraphics[width=0.85\linewidth]{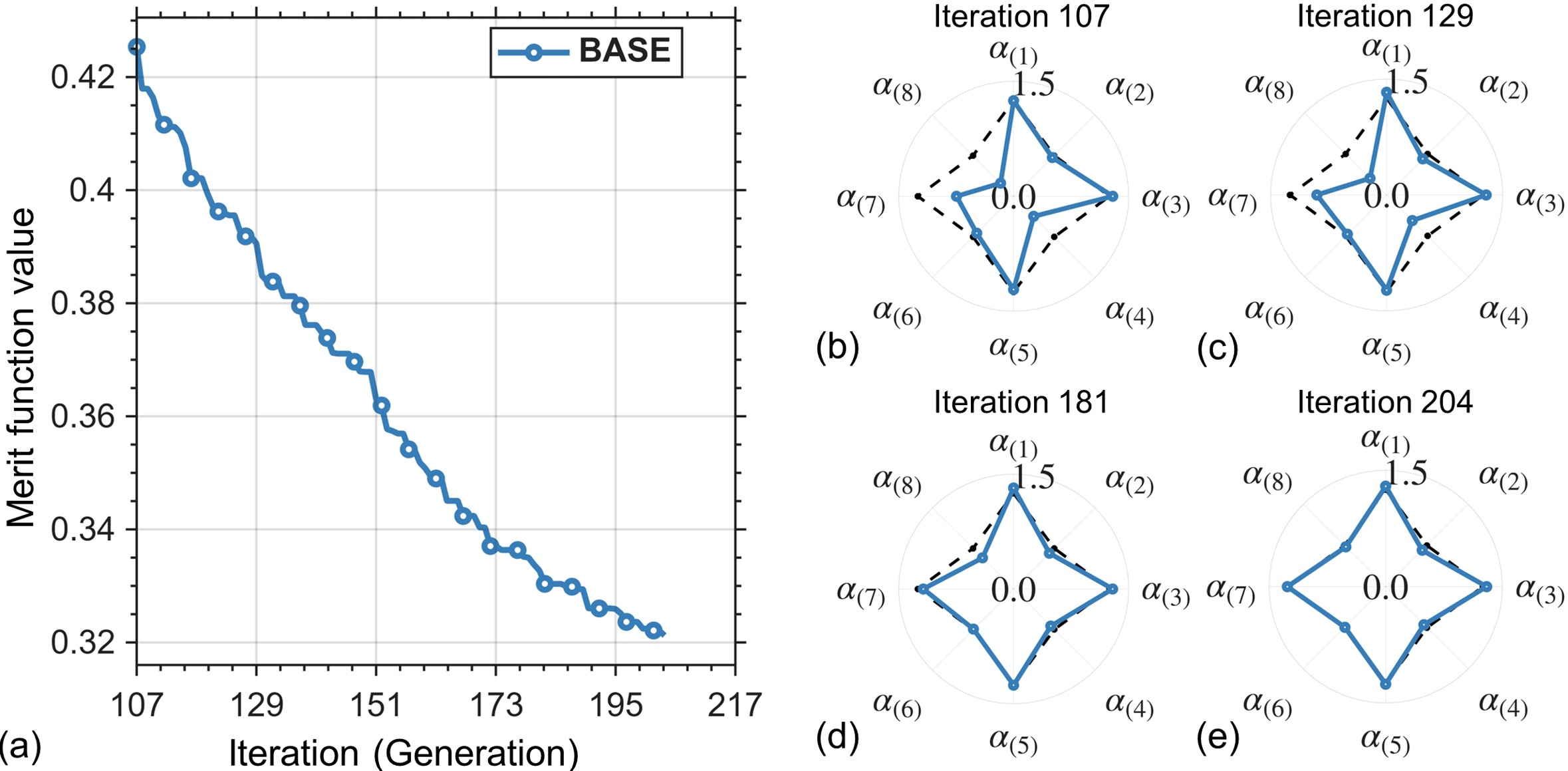}
	\caption{Iteration history of the merit function for the $d=8$ case using \textbf{BASE} (a), with the best designs at selected iterations shown on polar axes ((b)---(e)). The exact optimum $\bm{\alpha}_{\rm opt}$ is represented by the dotted line.}
	\label{fig:scalability_base}
\end{figure}

\begin{figure}[H]
	\centering
	\includegraphics[width=0.85\linewidth]{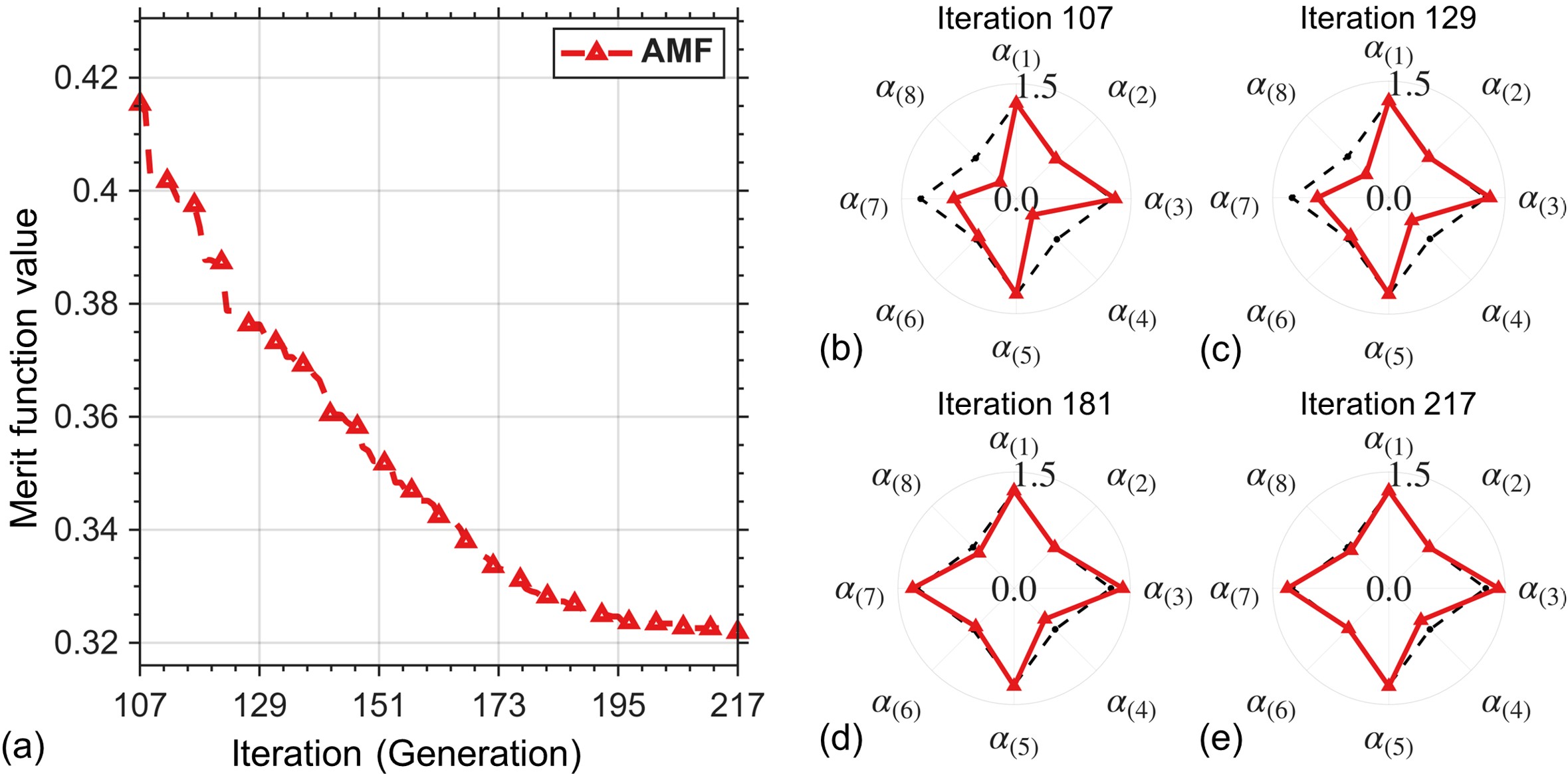}
	\caption{Iteration history of the merit function for the $d=8$ case using \textbf{AMF} (a), with the best designs at selected generations shown on polar axes ((b)---(e)). The exact optimum $\bm{\alpha}_{\rm opt}$ is represented by the dotted line.}
	\label{fig:scalability_amf}
\end{figure}

For example, Figures~\ref{fig:scalability_base} and \ref{fig:scalability_amf} compare the optimization histories for the $d=8$ case. The scaled merit function histories are similar, and both methods eventually satisfy the termination criterion of $2\%$ agreement with $c_{\rm opt}$. The best designs at selected iterations, shown on the polar axes, also gradually approach the analytical optimum. In this case, \textbf{AMF} requires $13$ additional iterations to reach the stopping criterion, which is a small difference compared with the approximately $200$ iterations overall.

\begin{figure}[H]
	\centering
	\includegraphics[width=0.6\linewidth]{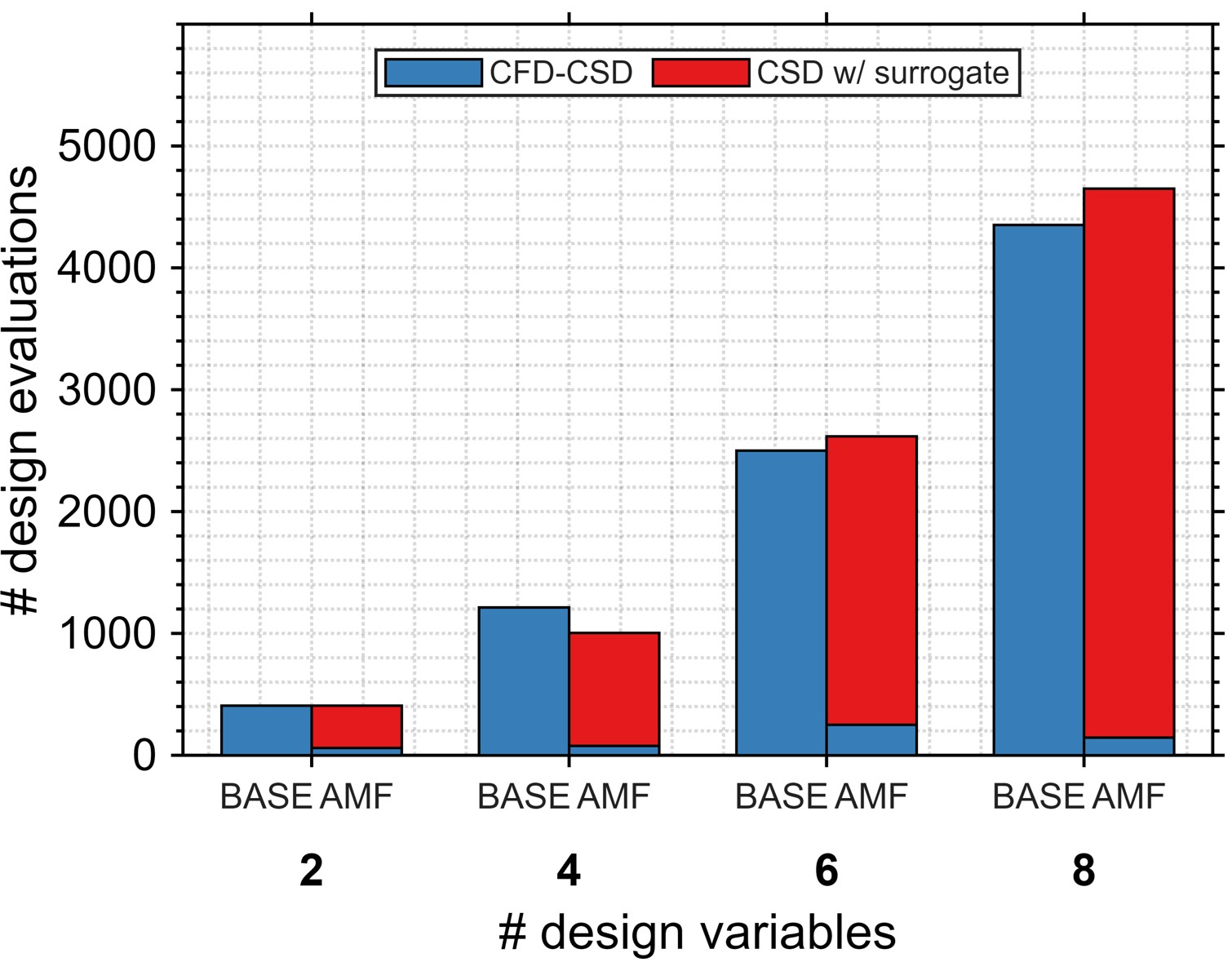}
	\caption{Scalability of the adaptive multi-fidelity framework (\textbf{AMF}) with respect to design-space dimension, compared with the baseline framework (\textbf{BASE}).}
	\label{fig:scalability}
\end{figure}

Figure~\ref{fig:scalability} shows the scaling of computational cost in terms of number of coupled and surrogate-assisted evaluations. The total number of design evaluations required to reach the termination threshold increases with $d$ and remains comparable between \textbf{BASE} and \textbf{AMF}. This indicates that, for the cases considered, using the surrogate fluid model does not substantially increase the number of design evaluations required by the genetic algorithm.

In \textbf{AMF}, coupled analyses account for $14.74\%$, $7.67\%$, $9.55\%$, and $3.11\%$ of all evaluations for $d=2$, $4$, $6$, and $8$, respectively. Because the computational cost of an evaluation is dominated by the fluid solver, the reduction in coupled analyses translates directly into significant computational savings. Reaching the stopping criterion requires $9.3\times10^3$, $2.7\times10^4$, $5.4\times10^4$, and $9.6\times10^4$ CPU core-hours with \textbf{BASE}, compared with $1.6\times10^3$, $2.9\times10^3$, $8.0\times10^3$, and $9.0\times10^3$ CPU core-hours with \textbf{AMF}. These values correspond to computational cost reductions of $82.8\%$, $89.3\%$, $85.2\%$, and $90.6\%$, respectively. Thus, \textbf{AMF} maintains a substantial reduction in computational cost as the design space dimension increases.

\subsection{Flexible panel under shock loading}

\subsubsection{Problem description}
\begin{figure}[H]
	\centering
	\includegraphics[width=0.9\linewidth]{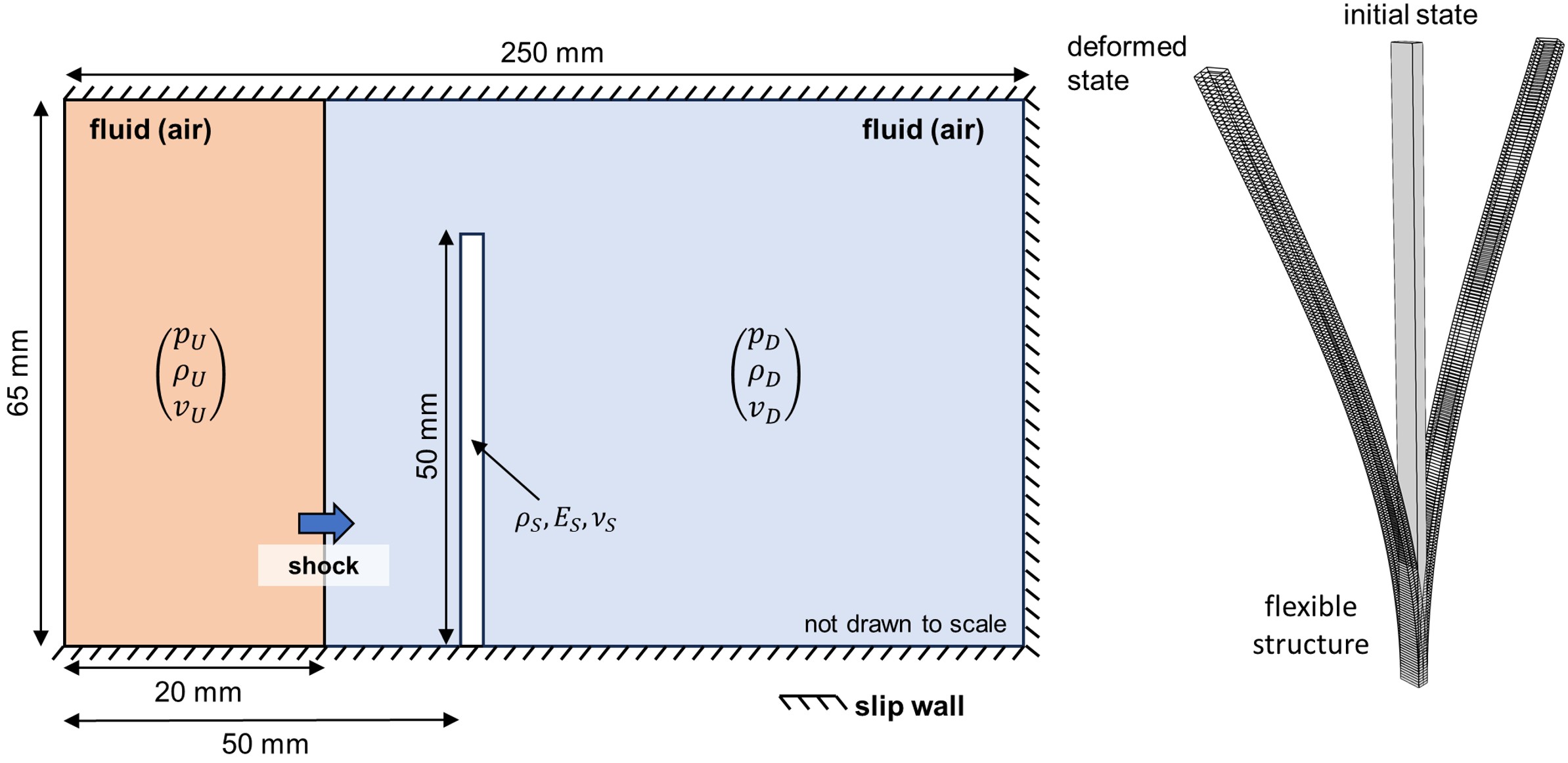}
	\caption{A flexible cantilever panel subject to shock loading.}
	\label{fig:numerical_model}
\end{figure}

Next, we demonstrate the proposed adaptive multi-fidelity optimization framework using the two-dimensional model problem illustrated in \figurename~\ref{fig:numerical_model}. The problem involves a slender panel subject to a planar shock wave. When the shock impinges on the panel, it produces a reflected wave traveling upstream, a transmitted shock that diffracts around the panel tip, and vortex structures that interact with reflected waves from the domain boundaries. These nonlinear, transient flow features generate a time-dependent pressure distribution on the panel surface. Meanwhile, the panel is designed to be flexible and bends under the fluid loading. 

This configuration was previously investigated by Giordano \textit{et al.}~\cite{giordano_shock_2005} as a canonical FSI problem relevant to a range of aerospace and defense applications. The geometric and material properties adopted in the present study largely follow their setup. The fluid flow surrounding the panel is modeled as a compressible, inviscid perfect gas governed by the Euler equations~\eqref{eq:euler_terms}. The computational domain  is a two-dimensional rectangular region of length $250$ mm and height $65$ mm. A planar shock wave with Mach number $1.21$ is initialized $20$ mm from the inlet boundary and propagates from left to right. The upstream fluid state is specified as $p_U = 155.7$ kPa, $\rho_U = 1.63$ kg\,m$^{-3}$, and $v_{U} = 109.6$ m\,s$^{-1}$, while the downstream state corresponds to ambient conditions with $p_D = 101$ kPa, $\rho_D = 1.2$ kg\,m$^{-3}$, and $v_{D} = 0$ m\,s$^{-1}$. The panel is $50$ mm tall and is made of steel, with density $\rho_S = 7600$ kg\,m$^{-3}$, Young's modulus $E_S = 220$ GPa, and Poisson's ratio $\nu_S = 0.33$. It is clamped at its base, while its other boundaries are free to deform and exposed to the fluid. Fluid-structure coupling is enforced through the kinematic and dynamic interface conditions defined in~\eqref{eq:kinematic_interf_cond} and~\eqref{eq:dynamic_interf_cond}, respectively. 

\subsubsection{Validation of coupled CFD--CSD analysis}\label{sec:validation_results}

\begin{figure}[H]
	\centering
	\includegraphics[width=\linewidth]{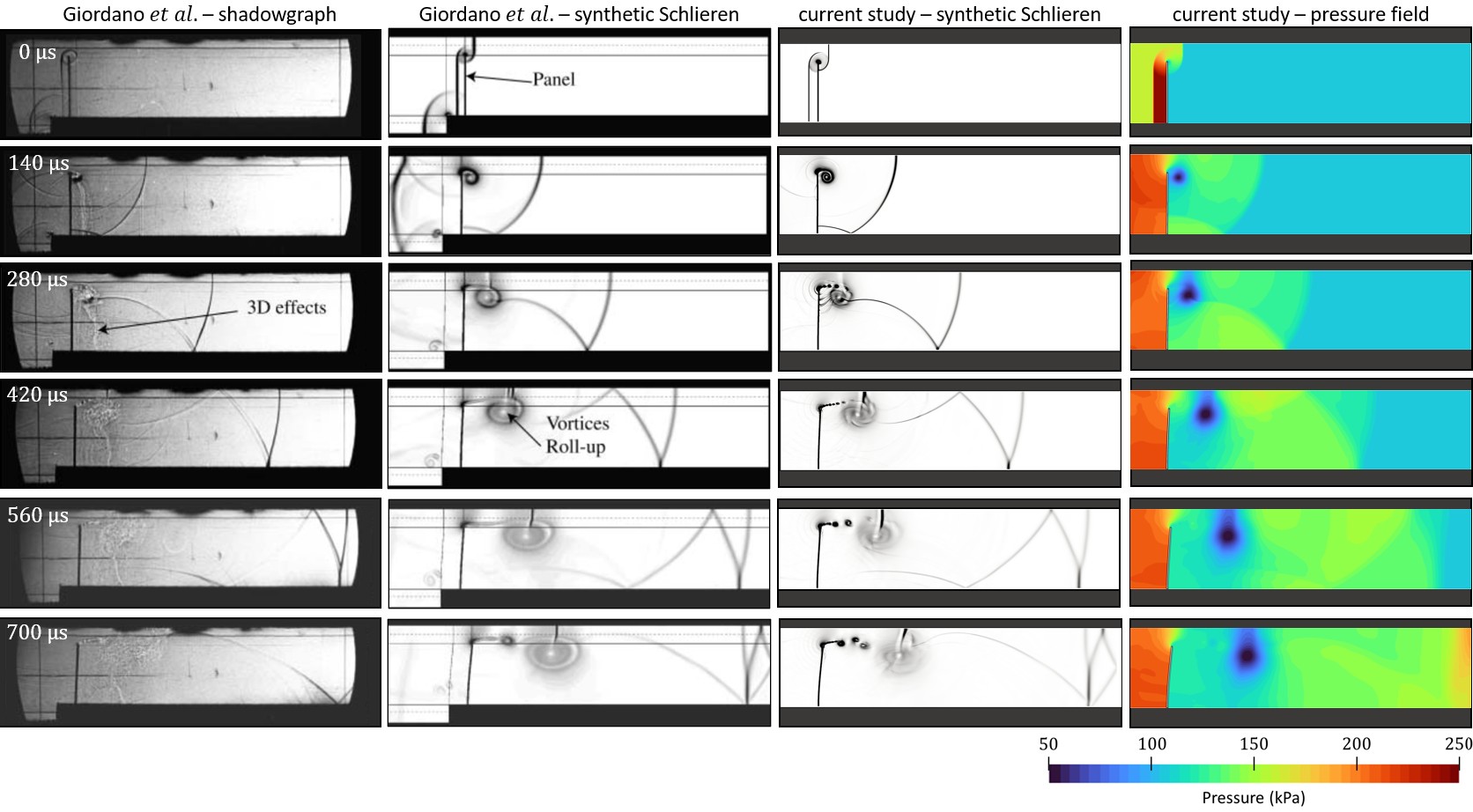}
	\caption{Comparison of current simulation result with published data for a reference design ($\bm{\alpha}=[1.0,~1.0,~1.0]^T$).}
	\label{fig:fluid_field_comp}
\end{figure}

For a reference panel with a uniform thickness of $1$ mm, we compare the results obtained using our simulation framework (i.e., M2C and Aero-S) with the numerical and experimental results from  Giordano \textit{et al.}~\cite{giordano_shock_2005}. The  fluid domain is discretized by a non-uniform, non-body conforming Cartesian mesh of $57,460$ nodes, with a minimum element size of $0.25$ mm near the fluid--structure interface. The panel is discretized with $0.5$ mm hexahedral elements, resulting in a total of $200$ elements.  \figurename~\ref{fig:fluid_field_comp} presents a qualitative comparison of the overall flow field using synthetic Schlieren images (i.e., the magnitude of density gradient) from the present study, together with the numerical Schlieren fields and experimental shadowgraph images extracted from Giordano \textit{et al.} When the incident planar shock reaches the flexible panel, the impact generates a reflected shock traveling upstream and a transmitted shock propagating downstream into the ambient fluid. As reported in the reference study, the transmitted shock subsequently develops into a curved front as the flow area increases behind the panel. As the flow evolves, vortex structures form near the panel tip and are convected downstream, interacting with other nonlinear flow features such as reflected shocks and expansion waves originating from the right wall. These backward-propagating waves eventually impact on the panel again, inducing additional deformation. Overall, the present results show good  agreement with both the experimental observations and the numerical results reported in the reference study.

In addition, quantitative comparison of the panel's tip displacement also shows reasonable agreement. The peak horizontal displacement predicted in the present simulation is $5.76~\text{mm}$, compared with the experimental measurement of $6.2\pm0.4~\text{mm}$ and the numerical result of $6.6~\text{mm}$ reported by Giordano \textit{et al.} Additional details can be found in~\cite{narkhede2026multifidelity}. It should be noted that the present simulation setup does not include the rigid ramp in the experimental and computational configurations of Giordano \textit{et al.}. This leads to some discrepancies in the result. Moreover, the simulation and experimental data reported in the reference study themselves exhibit noticeable differences, which are discussed in Giordano \textit{et al.}~\cite{giordano_shock_2005}.


\subsubsection{Design of optimization study}

We consider a shape optimization problem in which the panel geometry is modified by varying the thickness at three lengthwise sections. The design variable is defined as
$\boldsymbol{\alpha} = [\alpha_{(1)}, \alpha_{(2)}, \alpha_{(3)}]^T$, where $\alpha_{(1)}$, $\alpha_{(2)}$, and $\alpha_{(3)}$ denote the thicknesses (in mm) of the tip, mid-section, and clamped base, respectively. The thickness distribution between these locations is set by linear interpolation.

The optimization objective is to minimize the peak horizontal displacement of the panel's tip, $\norm{w_x(\bm{X}_{\text{tip}},t)}_\infty$, subject to constraints on the maximum von Mises stress at the clamped base $\sigma_{\mathrm{VM}}(\bm{X}_{\text{base}},t)$ and the total structural mass $M$. This constrained optimization problem is stated as
\begin{equation}
	\begin{aligned}
		\underset{\boldsymbol{\alpha} \in \Omega_D}{\operatorname*{minimize}} \quad
		&\norm{w_x(\bm{X}_{\text{tip}},t;\bm{\alpha})}_\infty\\
		\text{subject to} \quad
		& \sigma_{\mathrm{VM}}(\bm{X}_{\text{base}},t;\bm{\alpha})
		\leq 500~\text{MPa}, \\
		& M(\bm{\alpha}) \leq 0.76~\text{g},
	\end{aligned}
	\label{eq:opt_problem}
\end{equation}
where $\Omega_D = [0.5,\, 4.0]^3$ limits the design space to a bounded rectangular region.  The constrained problem is converted to an unconstrained one through an exterior penalty method, with the structural response normalized by reference values $w^{\text{ref}} = 6.0$ mm,
$\sigma^{\text{ref}} = 800$ MPa, and $M^{\text{ref}} = 0.76$ g.

In this problem, the fluid domain is substantially larger than the structural domain, meaning that the computational cost of each coupled analysis is dominated by the CFD solver. The objective and constraints are common in structural design optimization, and the trade-offs between stiffness and weight can be rationalized from elementary beam theory.

To evaluate the proposed framework, three optimization studies are conducted, differing only in how the fluid subproblem is treated.

\begin{itemize}
	\item \textbf{BASE}: The baseline optimization framework described in Sec.~\ref{sec:baseline_optimization} (i.e.,~SOFICS-Base) is employed. Because the CFD solver employs an embedded boundary method, the fluid mesh remains unchanged for all designs; only the structural mesh is updated. 
	
	\item \textbf{AMF}: The adaptive multi-fidelity framework described in Sec.~\ref{sec:adaptive_framework} (i.e.,~SOFICS-AMF) is employed. $\tau_e = 1.3\%$,  $\tau_u = 1\%$, and $N_b = 5$.
	
	\item \textbf{DRY}: Every design is evaluated by solving only the structural sub-problem $\mathcal{R}_\text{S}$. The fluid-induced loads are prescribed using the pressure time-history obtained from the coupled CFD-CSD analysis of the reference panel (i.e., $\bm{\alpha}=[1.0, 1.0, 1.0]^T$). This study can be viewed as a limiting case of \textbf{AMF} under large tolerances  $\tau_e$ and $\tau_u$. It also reflects the conventional structural optimization approach, in which external loads are specified {\it a priori}.
	
\end{itemize}

The design space is explored using a single-objective genetic algorithm. Table~\ref{tab:dakota_inputs} summarizes the main algorithmic settings and parameter values, most of which are default options in Dakota. The initial population is generated by Latin Hypercube Sampling (LHS) within the subdomain $[2.5,\,4.0]^3$ of the design space. This choice is deliberate. Panels in this region are uniformly thick, and therefore heavy and stiff. While the tip displacement is small and the stress constraint is easily satisfied, the mass constraint is violated, resulting in a high penalized merit function value for every initial design. The optimizer is thus forced to explore the design space incrementally towards the feasible region, thereby testing the surrogate model's adaptability as the evaluated designs shift progressively away from the initial cluster. Furthermore, the optimal design is expected to feature a tapered profile (i.e., $\alpha_{(1)} < \alpha_{(2)} < \alpha_{(3)}$), which lies far from the initial population. Starting from a distant, uniformly thick region therefore also demonstrates the ability of the genetic algorithm to locate the global minimum without being trapped in a local minimum.


\begin{table}[H]
	\centering
	\caption{Summary of Dakota input parameters used in the optimization study.}
	\label{tab:dakota_inputs}
	\begin{tabular}{l|l}
		\toprule
		\multicolumn{1}{c}{Item} & 
		\multicolumn{1}{c}{Inputs} \\
		\midrule
		
		\textbf{Seed}
		& $1649919949$ \\[2pt]
		
		\textbf{Initial population} 
		& Size: $50$ \\
		& Type: user-specified (randomly distributed within $[2.5,~4]^3$) \\ [2pt]
		
		\textbf{Fitness} 
		& Type: exterior-penalty merit function \\
		& Constraint penalty parameter: $20.0$ \\[2pt]
		
		\textbf{Crossover} 
		& Type: shuffle random \\
		& Number of parents: $2$ \\ 
		& Number of offspring: $2$ \\
		& Crossover rate: $0.8$ \\[2pt]
		
		\textbf{Mutation} 
		& Type: replace-uniform mutation \\
		& Mutation rate: $0.3$ \\[2pt]
		
		\textbf{Replacement} 
		& Type: elitist  \\[2pt]
		
		\textbf{Stop criterion} 
		& Maximum number of iterations: $30$ \\
		
		\bottomrule
	\end{tabular}
\end{table}


\subsubsection{Results and discussion}

\begin{figure}[H]
	\centering
	\includegraphics[width=1\linewidth]{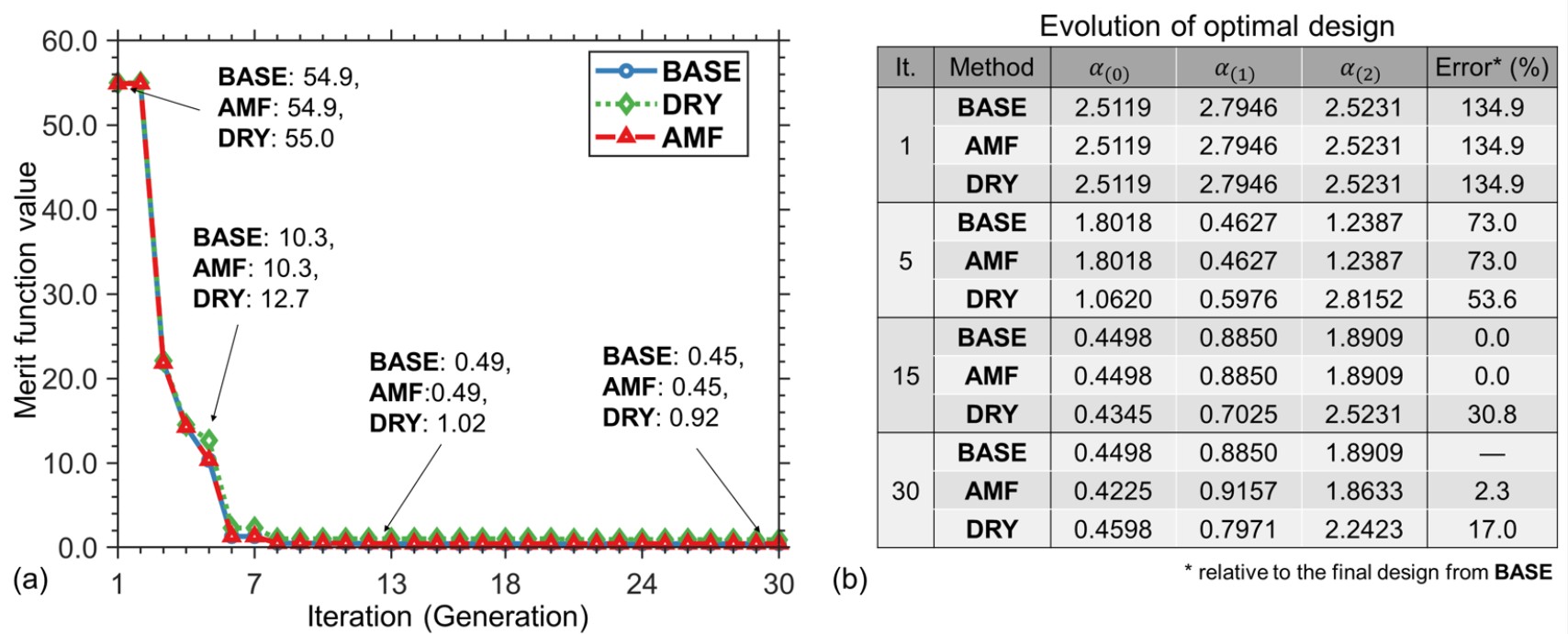}
	\caption{Comparison of the three methods \textbf{BASE}, \textbf{AMF}, and \textbf{DRY}. (a) Iteration history of the merit function; (b) evolution of the optimal design at four iterations. The error in $\bm{\alpha}$ is calculated using the final result from \text{BASE} as reference.}
	\label{fig:iteration_hist}
\end{figure}

In all three cases, the optimizer was run for $30$ iterations. As shown in \figurename~\ref{fig:iteration_hist}(a), the merit function value starts at approximately $55$ in each case. By the end of the optimization, all the three cases have yielded significant reduction in the merit function value. The value decreases to  $0.45$ for \textbf{BASE} and \textbf{AMF}, and to $0.92$ for \textbf{DRY}. \figurename~\ref{fig:iteration_hist}(b) shows the evolution of the optimal design throughout the optimization process. Taking the final result from \textbf{BASE} as the reference, the error in the \textbf{AMF} result is $2.3\%$, compared with $17.0\%$ for \textbf{DRY}.

\begin{figure}[H]
	\centering
	\includegraphics[width=1\linewidth]{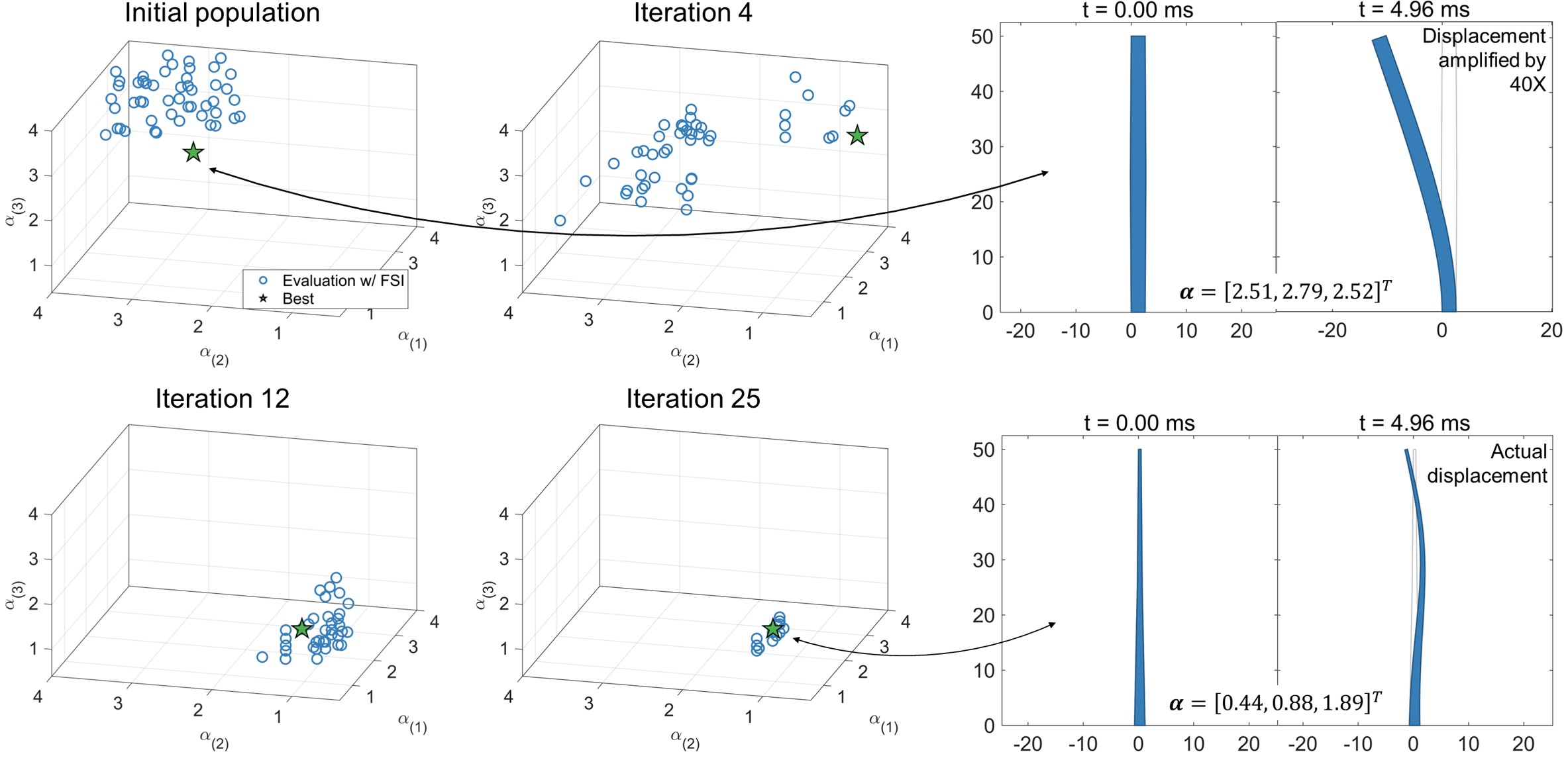}
\caption{Design population at selected iterations of the \textbf{BASE} optimization study. Scatter plots show candidate designs in the three-dimensional design space, with the current best design marked by a pentagram.}	\label{fig:generations_baseline}
\end{figure}

Figure~\ref{fig:generations_baseline} illustrates the evolution of the design population over successive iterations in \textbf{BASE}. The genetic algorithm (GA) initially focuses on exploring the design space, gradually shifting the population away from the initial samples with high panel thicknesses. By iteration $4$, the population begins to move toward lower-thickness regions, with a noticeable preference for designs in which $\alpha_{(2)}$ is relatively small, as reflected by the best design in that iteration. Continued exploration guides the GA toward a region containing tapered panel geometries around iteration $12$. This transition is also reflected in the merit function history shown in \figurename~\ref{fig:iteration_hist}, where the objective value improves by approximately one order of magnitude relative to the initial population. By iteration $25$, the population becomes tightly clustered, and the optimizer converges to an optimal design with thickness parameters $\displaystyle\bm{\alpha}_{\text{opt}}^{\text{(BASE)}} = [0.44,~0.88,~1.89]^T$ mm. The resulting configuration exhibits a pronounced taper from tip to base, consistent with the physical expectation that the required structural stiffness increases toward the clamped support, where bending moments are largest.

\begin{figure}[H]
\centering
\includegraphics[width=1\linewidth]{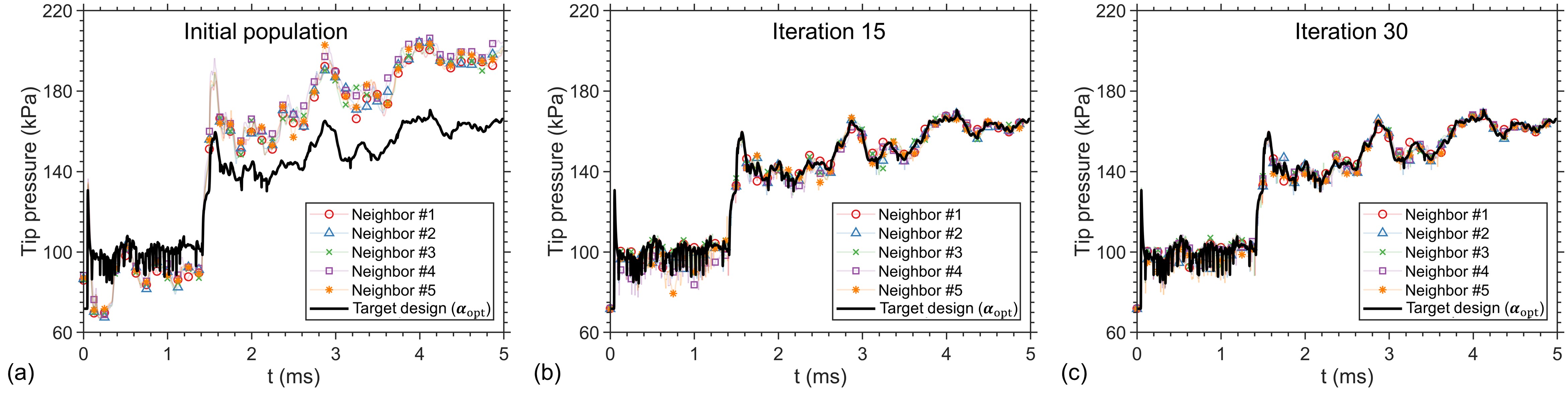}
\caption{Pressure time histories recorded at the panel's tip across selected iterations of the \textbf{BASE} optimization study.}
\label{fig:pressure_history}
\end{figure}

The convergence behavior observed from \textbf{BASE} provides a rationale for the proposed adaptive multi-fidelity framework. Figure~\ref{fig:pressure_history} shows the pressure time history at the panel tip. The result corresponding to the optimal design $\displaystyle\bm{\alpha}_{\text{opt}}^{\text{(BASE)}}$ is shown as a thick solid line. Three iterations---1, 15, and 30---are selected for comparison. For each iteration, the five nearest neighbors of $\displaystyle\bm{\alpha}_{\text{opt}}^{\text{(BASE)}}$ among the evaluated designs up to that iteration are identified, representing the designs that would be used for interpolating the fluid-induced load at $\displaystyle\bm{\alpha}^{\text{(BASE)}}_{\text{opt}}$ in the surrogate model. The pressure histories associated with these neighboring designs are plotted in different colors.

In the initial population, the sampled designs lie far from $\displaystyle\bm{\alpha}_{\text{opt}}^{\text{(BASE)}}$ (see \figurename~\ref{fig:generations_baseline}), and the pressure histories at the five neighbors deviate substantially from that of $\displaystyle\bm{\alpha}_{\text{opt}}^{\text{(BASE)}}$. This reflects the FSI in the problem: the pressure loading on the structure varies with the structural design variables. It also implies the necessity of a decision model that limits the use of surrogate evaluations in such situations, reverting to coupled CFD--CSD analyses until sufficient local data have been accumulated. In later iterations (e.g., 15 and 30), as the population concentrates within a small neighborhood of the optimal design, the pressure histories associated with neighboring designs converge toward a common profile. As a result, the accuracy of the interpolation surrogate model improves. The GPR-based decision model is expected to track this evolving accuracy and progressively increase the use of surrogate evaluations throughout the optimization.
\vspace{2mm}

\begin{figure}[H]
	\centering
	\includegraphics[width=0.95\linewidth]{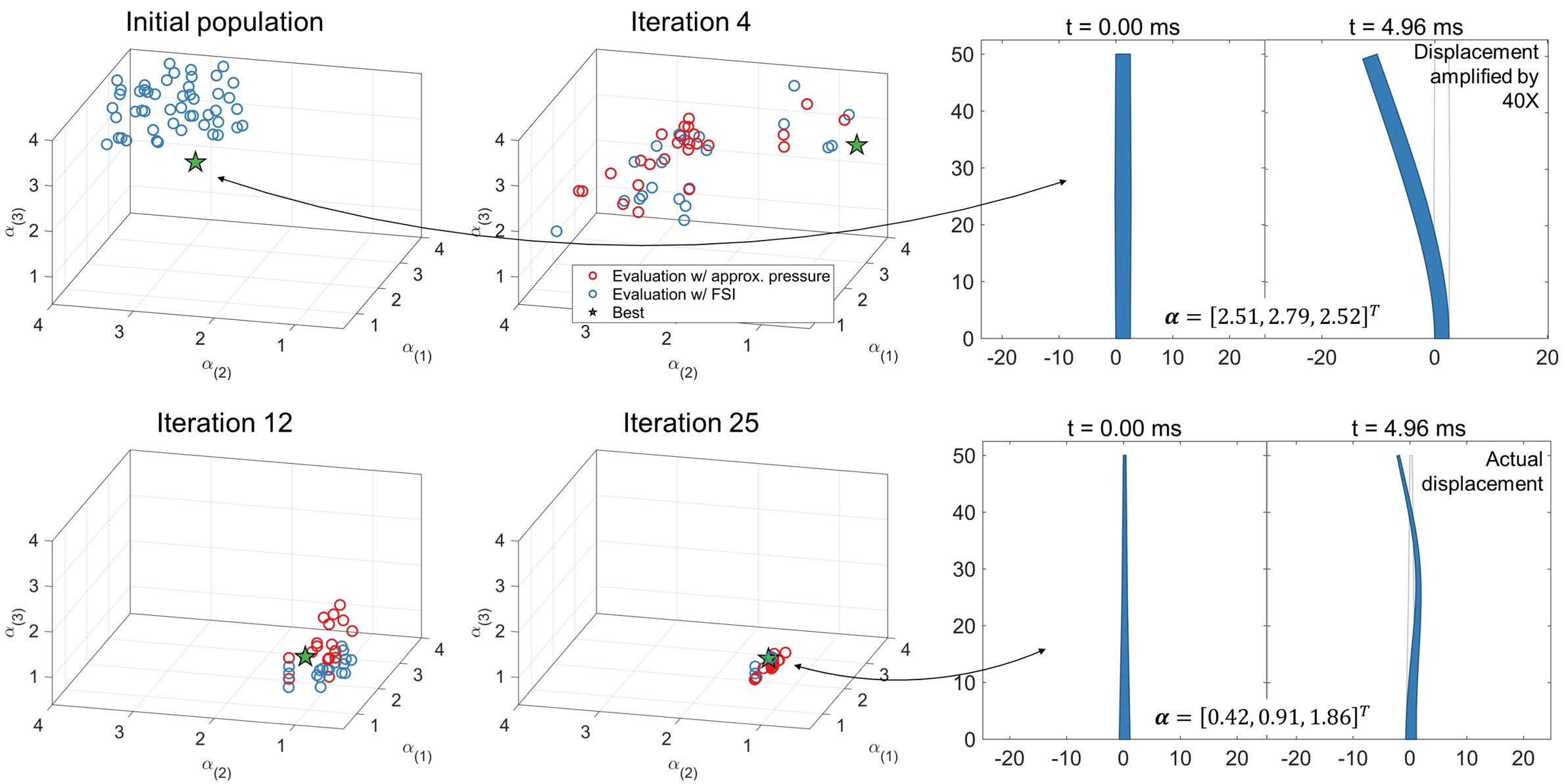}
	\caption{Evolution of the design population in the \textbf{AMF} optimization study. Blue and red circles indicate design candidates evaluated using coupled simulation and surrogate fluid model, respectively. The best design in each iteration is marked by a pentagram.}
	\label{fig:generations_adaptive}
\end{figure}

Figure~\ref{fig:generations_adaptive} presents the results obtained from \textbf{AMF} at the same iterations as shown in \figurename~\ref{fig:generations_baseline}.   Design candidates evaluated with the surrogate model are shown in red color. The adaptive multi-fidelity framework reproduces the convergence behavior of \textbf{BASE} while substantially reducing the number of coupled CFD--CSD analyses. In the initial iteration, all $50$ candidate designs are evaluated using the coupled FSI solver. This number decreases to $8$ in iteration 4, $12$ in iteration 12, and $1$ in iteration 25. 

The number of surrogate-accepted designs is influenced by the crossover and mutation operators and their prescribed rates (Table~\ref{tab:dakota_inputs}). With a crossover rate of $0.8$, most new design points inherit combinations of variables from two parent designs. As a result, they tend to remain within the convex hull of the current population, where the GPR model has already accumulated data and its predictive uncertainty is low. The mutation operator is applied with a rate of $0.3$. It generates new designs by replacing one or more variables with values drawn uniformly from the feasible range. Consequently, some designs are placed in unexplored regions of the design space. In these regions, the GPR uncertainty is more likely to exceed the prescribed threshold $\tau_u$, triggering the use of coupled CFD--CSD analysis. As a result, even in later iterations when surrogate-assisted evaluations dominate, a small but consistent number of designs still require coupled analysis. These cases typically correspond to mutated offspring that explore sparsely sampled regions of the design space.

\begin{figure}[H]
	\centering
	\includegraphics[width=1\linewidth]{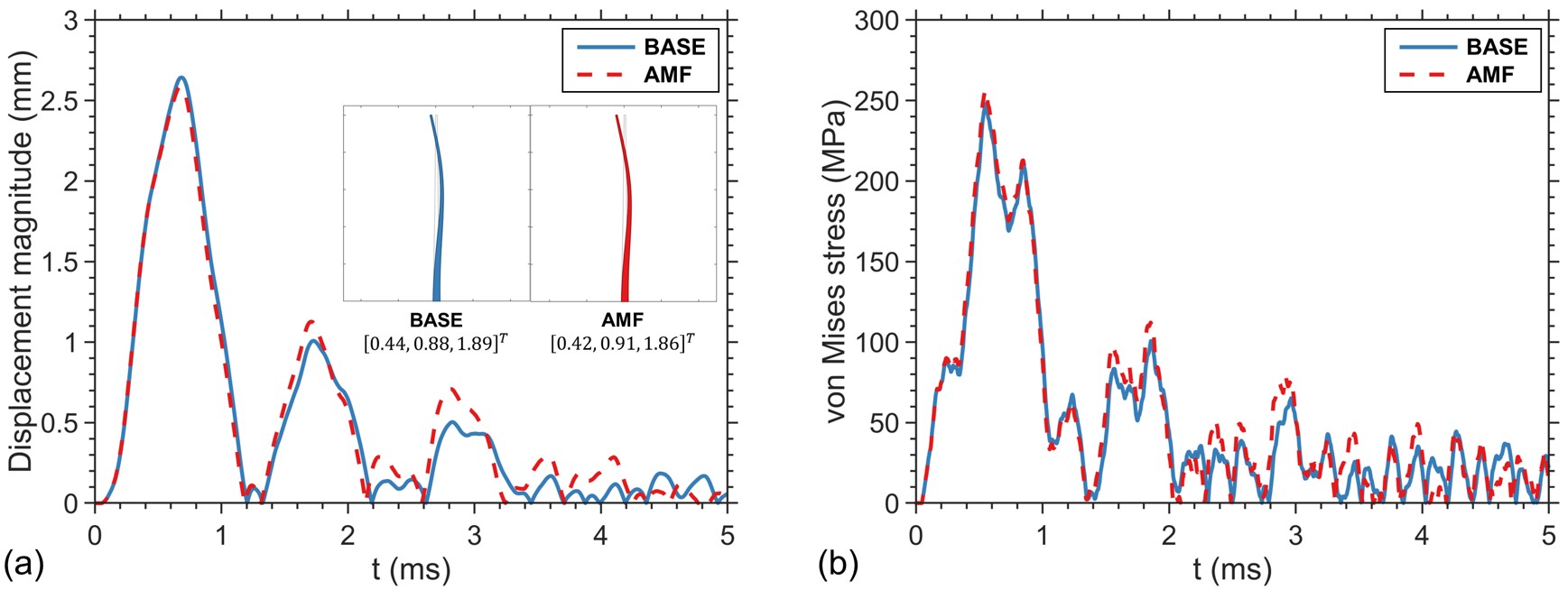}
	\caption{Comparison of the dynamic responses of the optimal designs obtained from \textbf{BASE} and \textbf{AMF}. (a) Time history of the panel's tip displacement, and (b) time history of the von Mises stress at the panel's clamped base.}
	\label{fig:disp_compare}
\end{figure}

The final design obtained from \textbf{AMF} is $\displaystyle\bm{\alpha}_{\text{opt}}^{\text{(AMF)}} = [0.42, 0.91, 1.86]^T$. It differs from the result of \textbf{BASE} by only $2.3\%$. Fig.~\ref{fig:disp_compare} compares the dynamic responses of these two designs in coupled CFD-CSD analyses. The designs show very close agreement, both in geometry and in their transient responses during the FSI process. In contrast, the result obtained from \textbf{DRY} is $\displaystyle\bm{\alpha}_{\text{opt}}^{\text{(DRY)}} = [0.46, 0.80, 2.24]^T$, which differs from the \textbf{BASE} result by $17.0\%$. 

Comparing Figs.~\ref{fig:generations_baseline} and \ref{fig:generations_adaptive}, it is notable that \textbf{AMF} matches \textbf{BASE} not only in the final design but also in the evolution of the population during the optimization process. At each iteration, many designs in the current population closely correspond between the two runs. This suggests that from the optimizer's perspective, \textbf{AMF} and \textbf{BASE} are largely interchangeable. Meanwhile, the discrepancy between \textbf{DRY} and \textbf{BASE} is reasonable. Because the fluid-induced loads are pre-specified and fixed across all designs, \textbf{DRY} essentially solves a different optimization problem.

\begin{figure}[H]
	\centering
	\includegraphics[width=0.9\linewidth]{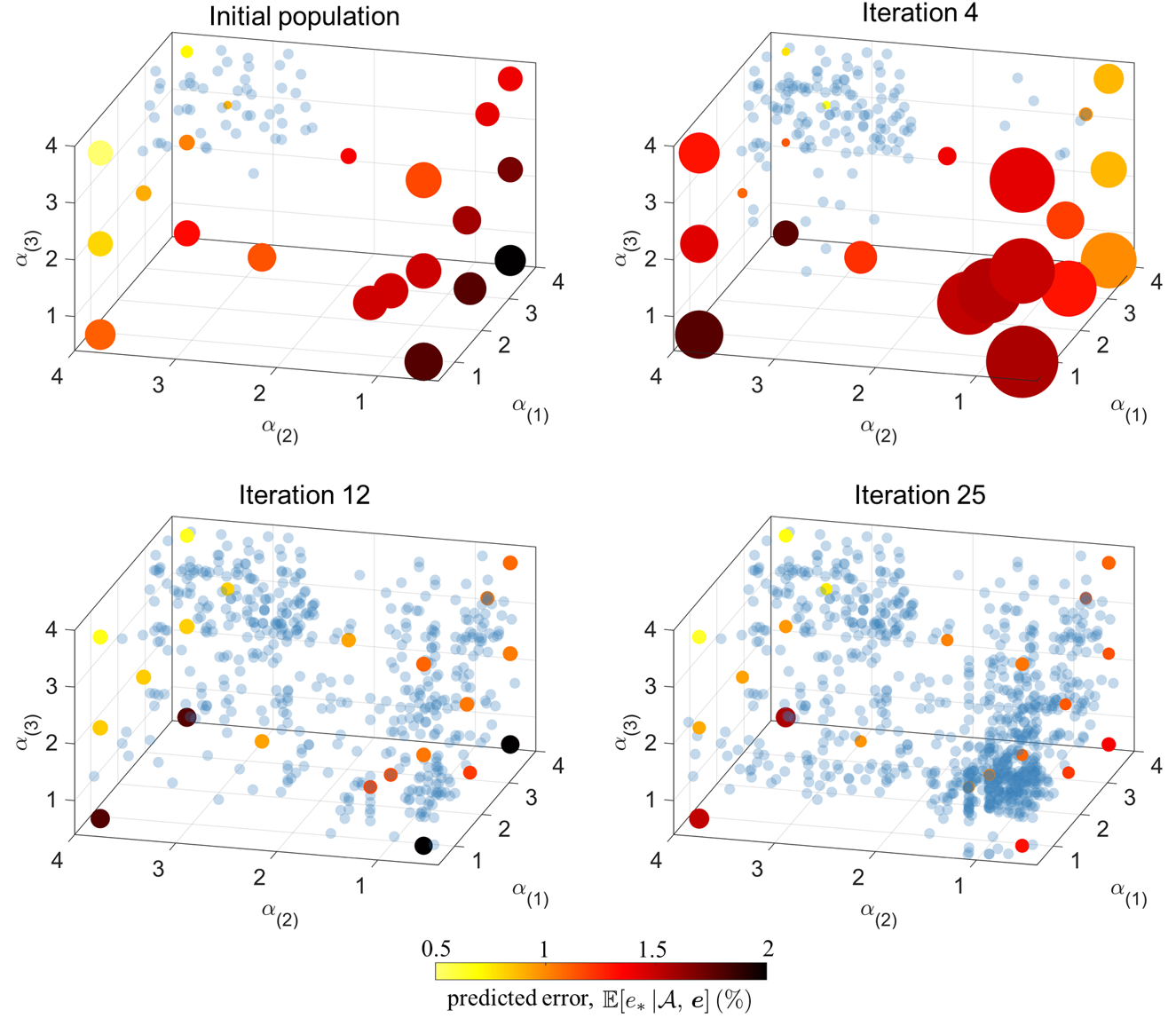}
\caption{Surrogate error and uncertainty at $21$ points in the design space as estimated by the GPR model, based on CFD-CSD simulation data accumulated up to Iterations 1, 4, 12, and 25. A filled circle is plotted at each point, with color and size representing the mean error $\mathbb{E}\big[e_* \mid \mathcal{A}, \bm{e}\big]$ and the associated uncertainty $u_*$, respectively. For reference, the available data points (i.e., $~\mathcal{A}$) used for both the surrogate model and the GPR model are shown as transparent blue circles.}	\label{fig:err_history}
\end{figure}

Next, we examine the performance of the GPR model for surrogate error estimation. Figure~\ref{fig:err_history} presents the predicted mean error and associated uncertainty at $21$ points in the rectangular design space, including eight points near the corners, eight points along the boundary of the mid-surface in $\alpha_{(3)}$, and five interior points corresponding to the optimal designs obtained in five iterations. Predictions based on data accumulated up to iterations 1, 4, 12, and 25 are shown separately. The results indicate that both the predicted error and uncertainty are generally smaller at sample locations lying within clusters of available data. Accordingly, as the iterative process advances and additional samples are collected, the overall prediction error and uncertainty decrease. It is noteworthy that error and uncertainty are not always proportional to each other. There are points in the design space, particularly during the early iterations, where the predicted error is small, yet the associated uncertainty is large. This observation supports the use of both error and uncertainty in the decision model. Moreover, even toward the end of the optimization, some points far from the optimal design still exhibit relatively large surrogate errors. This behavior reflects a deliberate aspect of our adaptive strategy: the goal is not to construct a globally accurate surrogate across the entire design space, which is unnecessary, but rather to ensure sufficient accuracy in regions where the optimizer explores.

\begin{figure}[H]
\centering
\includegraphics[width=0.6\linewidth]{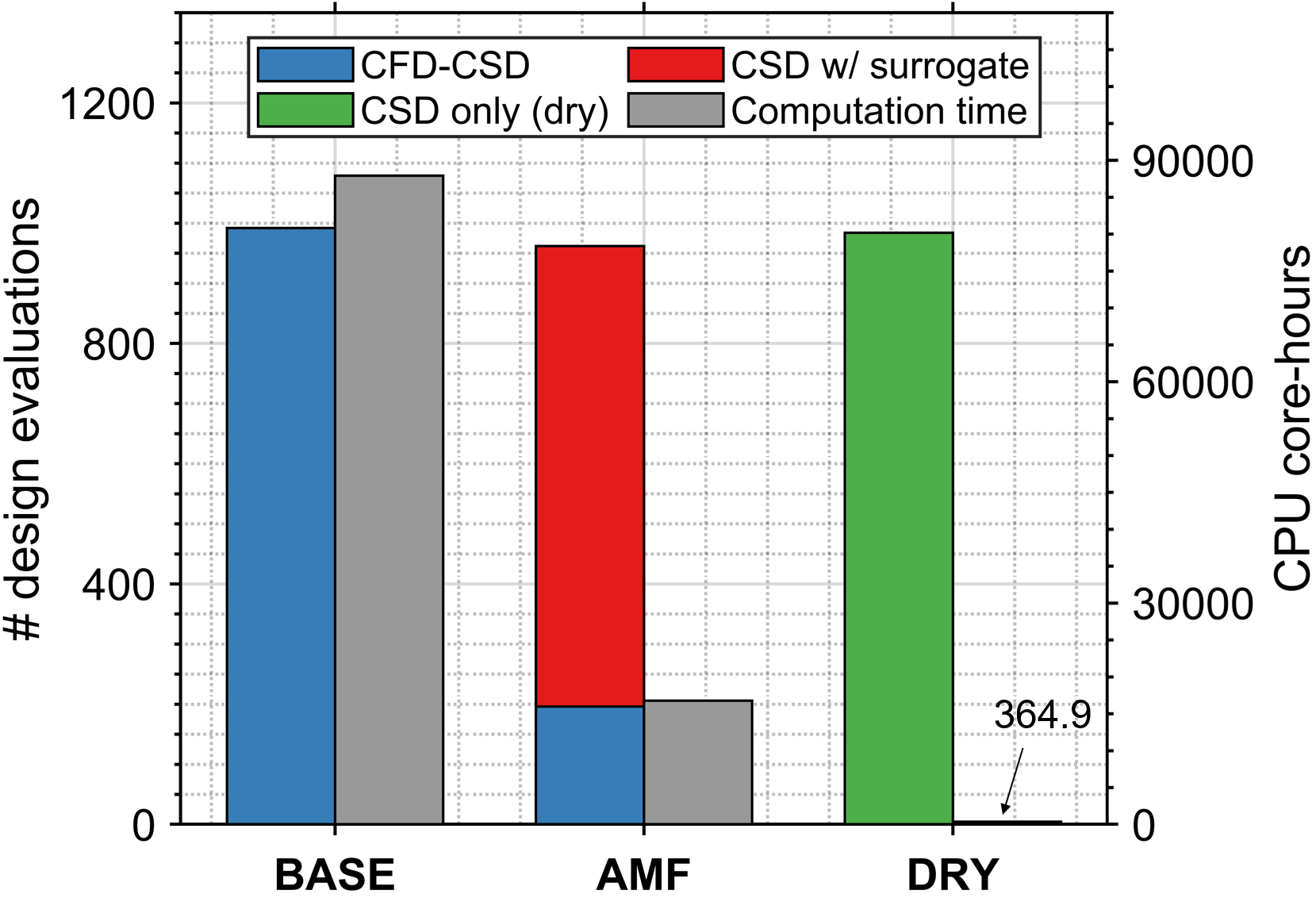}
\caption{Cost comparison for the three studies \textbf{BASE}, \textbf{DRY}, and \textbf{AMF}. The number of design evaluations is shown on the left axis; the total computation time is shown on the right axis.}
\label{fig:cost}
\end{figure}

Lastly, we compare the  computational cost of the three studies. As shown in \figurename~\ref{fig:cost}, \textbf{BASE} required $992$ coupled CFD-CSD analyses, resulting in a total cost of approximately $8.8\times 10^4$ CPU core-hours (on the TinkerCliffs computing cluster at Virginia Tech; same below). In comparison, \textbf{AMF} reduced both the number of coupled analyses and the total computational cost by about $80\%$, to $196$ analyses and $1.7\times 10^4$ core-hours, respectively. Although an additional $766$ CSD simulations using the surrogate fluid-load model were performed, their cost was negligible relative to the coupled analyses. \textbf{DRY}, by contrast, required $982$ CSD simulations and was indeed computationally inexpensive, with a total cost of only $364.9$ core-hours. However, it produced a design with a substantially larger discrepancy ($17.0\%$).

\section{Concluding remarks}\label{sec:conclusion}

New high-fidelity analysis capabilities do not directly translate into new design optimization capabilities, as their computational cost is often too high for many-query applications. This is presently the case for structures and vehicles operating under FSI. In practice, designers can often afford optimization based on CSD analyses alone, but not the sharp, orders-of-magnitude increase in cost associated with coupled CFD--CSD analyses.

In this paper, we have presented an adaptive multi-fidelity optimization framework tailored to such problems. The framework builds on recent progress in CFD--CSD coupling, interface tracking, and improved automation of mesh generation. It retains high-fidelity CSD analysis throughout the optimization process. For the computation of fluid-induced loads, it alternates between CFD and a data-driven surrogate, with the choice guided by a GPR model that estimates surrogate error together with uncertainty. The framework does not require any pre-computation such as dataset generation or model training. It is compatible, through parameter adjustments, with both more expensive optimization strategies that rely entirely on coupled analysis and inexpensive structure-only optimization using a fixed load time history. Although not explored directly in this paper, previously available coupled analysis data can be incorporated into the framework by initializing $\mathcal{D}$ and $\mathcal{A}$ with those data at the start of the algorithm.

This adaptive multi-fidelity framework addresses the fundamental trade-off between computational cost and predictive accuracy in design optimization. In this work, we have supported the proposed approach both theoretically, using a simplified FSI model (Sec.~\ref{sec:motivation}), and numerically through two benchmark problems (Sec.~\ref{sec:results}). In particular, the flexible panel under shock loading case demonstrates that the adaptive strategy reduces computational cost by $80\%$, while introducing only a $2.3\%$ error in the final design.


The proposed framework is modular in nature, as it builds on a partitioned procedure with separate  CFD and CSD solvers for the fluid and structural sub-problems. This modularity allows straightforward extension to alternative fluid and structural models, solvers, surrogate  models, and model management strategies. For example, the unstructured local interpolation adopted in this work provides a lightweight surrogate that can be efficiently updated as new  CFD data become available. However, the present study does not establish this interpolation approach as optimal. Alternative methods, including projection-based ROMs and various machine learning methods, can be  incorporated into the framework and warrant investigation in future work. Furthermore, the present study assumes that the optimization objective and constraints do not explicitly depend on the fluid state variables. It also considers a deterministic design problem, with uncertainty introduced only in assessing the fidelity of the fluid-load surrogate. Relaxing the former assumption would allow objectives and constraints involving fluid quantities to be considered directly, while extending the latter to account for uncertainties in the physical model, operating conditions, or design parameters would enable robust or reliability-based design optimization. Both represent natural directions for extending the range of problems addressed by the framework.



\section*{Acknowledgments}
	
The authors gratefully acknowledge support from the Office of Naval Research (ONR) under Award Nos.~N00014-23-1-2447 and N00014-24-1-2509; the Naval Air Warfare Center Aircraft Division under Contract No.~N6833525C0444; the National Institutes of Health (NIH) under Award Nos.~2R01-DK052985-26 and R01-DK138972; and a research contract from DynaSafe US LLC to Virginia Tech.

\bibliography{references}

\end{document}